\documentclass[aps,pre,twocolumn,superscriptaddress,amsmath,amssymb]{revtex4-1}
\usepackage{graphicx}
\usepackage{float}
\usepackage{bm} 
\usepackage{verbatim}  
\usepackage{mathrsfs}
\usepackage{xcolor}

\newcommand{\vect}[1]{\boldsymbol{\mathbf{#1}}}
\newcommand{\red}{\color{black}}

\usepackage{dcolumn}  
\usepackage[caption=false]{subfig}

\usepackage{diagbox}
\usepackage{array}
\newcolumntype{C}{>{$}c<{$}} 

\begin{document}

\title{Kinetic theory of pattern formation in mixtures of microtubules and molecular motors}

\author{Ivan Maryshev}
\affiliation{The University of Edinburgh, School of Biological Sciences, Institute of Cell Biology, Centre for Synthetic and Systems Biology,
Max Born Crescent, Edinburgh, EH9 3BF, United Kingdom}

\author{Davide Marenduzzo}
\affiliation{SUPA, School of Physics and Astronomy, The University of Edinburgh, James Clerk Maxwell Building, Peter Guthrie Tait Road, Edinburgh, EH9 3FD, United Kingdom}

\author{Andrew B. Goryachev}
\affiliation{The University of Edinburgh, School of Biological Sciences, Institute of Cell Biology, Centre for Synthetic and Systems Biology,
Max Born Crescent, Edinburgh, EH9 3BF, United Kingdom}

\author{Alexander Morozov}
\email{alexander.morozov@ph.ed.ac.uk}
\affiliation{SUPA, School of Physics and Astronomy, The University of Edinburgh, James Clerk Maxwell Building, Peter Guthrie Tait Road, Edinburgh, EH9 3FD, United Kingdom}

\date{\today}
\begin{abstract}
In this study we formulate a theoretical approach, based on a Boltzmann-like kinetic equation, to describe pattern formation in two-dimensional mixtures of microtubular filaments and molecular motors. Following the previous work by Aranson and Tsimring [Phys. Rev. E {\bf 74}, 031915 (2006)] we model the motor-induced reorientation of microtubules as collision rules, and devise a semi-analytical method to calculate the corresponding interaction integrals. This procedure yields an infinite hierarchy of kinetic equations that we terminate by employing a well-established closure strategy, developed in the pattern-formation community and based on a power-counting argument. We thus arrive at a closed set of coupled equations for slowly varying local density and orientation of the microtubules, and study its behaviour by performing a linear stability analysis and direct numerical simulations. By comparing our method with the work of Aranson and Tsimring, we assess the validity of the assumptions required to derive their and our theories. We demonstrate that our approximation-free evaluation of the interaction integrals and our choice of a systematic closure strategy result in a rather different dynamical behaviour than was previously reported. Based on our theory, we discuss the ensuing phase diagram and the patterns observed.
\end{abstract}

\maketitle

\section{Introduction}

Self-organisation of mixtures of biological polymers and molecular motors provides a fascinating manifestation of active matter~\cite{Marchetti2013,Needleman2017}. Microtubules are actively re-oriented by the molecular motors, and can form far-from-equilibrium global, cell-scale structures, such as the mitotic spindle apparatus~\cite{Mogilner2010}. It is believed that different motor types favour formation of distinct patterns: microtubule-sliding motors organise antiparallel bundles, while clustering motors control the formation of spindle poles and asters~\cite{Mogilner2010,Burbank2007}. 

Despite steady advance in the experimental analysis of such systems~\cite{Nedelec1997,Surrey2001,Hentrich2010,Sanchez2011,Sanchez2012,Foster2015,Brugues2014,Dogterom2013}, their theoretical description is stymied by the paucity of approaches able to connect individual microscopic motor-induced interactions of filaments to the macroscopic dynamics at lengthscales relevant to the whole cytoskeleton. Here we build on a kinetic method developed earlier in~\cite{Aranson2005,Aranson2006} to provide a revised version of the hydrodynamic equations that govern collective behaviour of microtubules in the presence of clustering motors. 

Microtubules are long and stiff rod-like biopolymers \cite{howard2001mechanics}. Because of the asymmetry of the constituting tubulin subunits, the microtubule filament has intrinsic orientation and distinct ends denoted as `-' and `+'. Molecular motors use chemical energy stored as ATP to move processively along microtubule filaments in one preferred direction. Some motors can bind two filaments simultaneously and, therefore, reorient and translocate them with respect to each other~\cite{howard2001mechanics,Cross2014}. The activity of such bivalent motors generates global order on a scale which is much larger than the length of a single filament.

Spontaneous transitions to various ordered states and patterns has been extensively studied in several \textit {in vitro} experiments with cell extracts~\cite{Nguyen2014,Brugues2014} and in the reconstituted systems containing mixtures of stabilised microtubules and purified motors. The latter systems recapitulate formation of structures with nematic~\cite{Sanchez2011,Sanchez2012} or polar order, e.g., asters and vortices~\cite{Nedelec1997,Surrey2001,Hentrich2010}. This type of systems is considered further in the current contribution.

Multiple approaches had been developed to advance our understanding of the dynamics typical of microtubule-motor mixtures. Besides direct agent-based simulations~\cite{Surrey2001,Loughlin2010}, mean field equations have been derived first on the basis of symmetry considerations~\cite{YounLee2001}, and then from the detailed microscopic rules of interaction \cite{Aranson2006,Liverpool2003,Ahmadi2006,Shelley2016}.

In the kinetic approach employed by Aranson and Tsimring \cite{Aranson2005,Aranson2006,Ziebert2007}, pairwise motor-mediated interactions of microtubules were treated as instantaneous collisions. These authors considered plus-directed clustering motors, which can align and bundle microtubules. Hydrodynamic equations for the two field variables, filament concentration and orientation, were derived by coarse-graining of the corresponding Boltzmann-type equation for the probability distribution function (PDF). Their model successfully recapitulated such phenomena as spontaneous ordering, bundling and formation of asters and vortices.

In this paper we revisit this technique. We demonstrate that using the exact form of the collision rate function, instead of the phenomenological expression suggested in~\cite{Aranson2006}, yields the system of equations with significantly distinct ``phase diagram''. Specifically, we find that in our model instabilities occur in a different order. We also argue that the previously neglected excluded volume effect needs to be considered to prevent density blow-up in the bundling regime. Additionally, we compare the closure of the equation expansion considered in~\cite{Aranson2006} with the more conventional method, which is historically used in Landau-Ginzburg-like equations~\cite{CrossHohenberg}. The latter approach necessitates introduction of an additional variable, the nematic order parameter or Q-tensor. The introduction of this additional field variable results in a novel instability, not observed in the previous work. 

The paper is organised as follows. Our kinetic model is introduced in Section II. In Section III we rederive the hydrodynamic equations obtained in~\cite{Aranson2006} and critically discuss the approximations used in this derivation. We demonstrate that the model exhibits bundling instability and argue for the need to include excluded volume effects. Section IV presents the derivation of such excluded volume terms.  In Section V we discuss evaluation of the interaction integrals. Our equations of motion are presented in Section VI: these are derived according to two different types of closure. We perform stability analysis of the equations of motion corresponding to both these closure schemes, present the corresponding phase diagrams, and provide the results of numerical simulations in Section VII. Finally, Section VIII contains a discussion of our results.

\section{Kinetic theory}
\label{sec:kineticmodel}

\subsubsection{The Boltzmann-like kinetic equation}

The setup of our problem follows that of~\cite{Aranson2006}. We consider a two-dimensional collection of microtubules, which we treat as slender rigid rods of length $L$. Since microtubules are polar objects, we describe their local orientation by a vector $\mathbf{n}$ that points from the 'minus'- to the 'plus'-end of a microtubule. We introduce a Cartesian coordinate system $(x,y)$, and parametrise the orientation vector by a single angle -- i.e., $\mathbf{n}=\left(\cos\phi,\sin\phi\right)$.
To describe spatial and orientational inhomogeneities in the system, we introduce the probability distribution function $P(\mathbf{r},\phi,t)$, defined in the usual way: $P(\mathbf{r},\phi,t)d\mathbf{r}d\phi$ gives the number of microtubules in a small volume of the phase space $d\mathbf{r}d\phi$ which are at position $\mathbf{r}$ and possess an orientation given by $\phi$ at time $t$. 

\begin{figure}[H]
\centering
\includegraphics[width=0.25\textwidth]{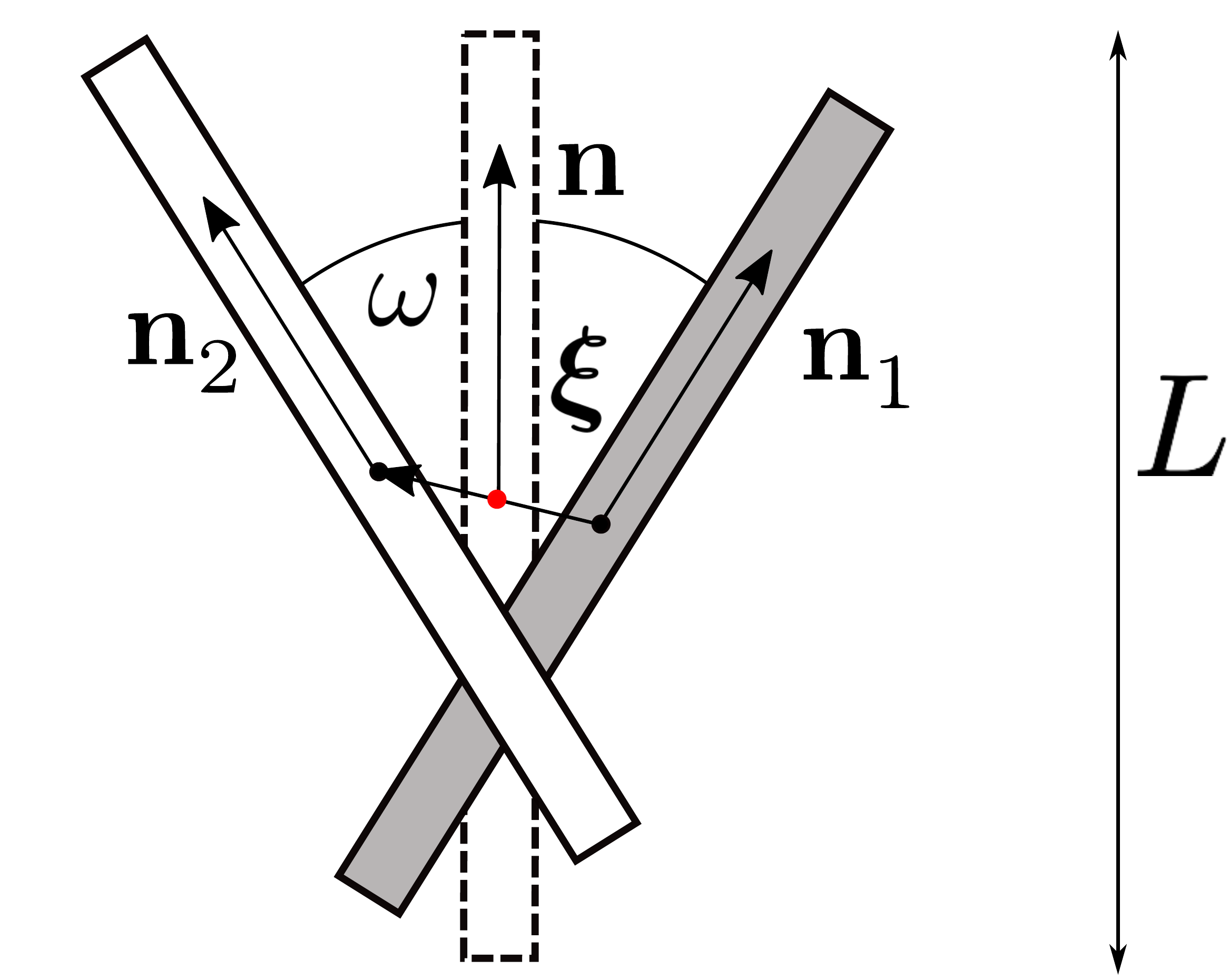}
\caption{Collision rule employed in Eq.\eqref{MEfull}. Two colliding microbutular bundles are re-oriented by the action of the molecular motors to assume a common orientation along the bisector of the original angle between them. Their centre of mass does not move in the process.}
\label{fig:Collision_rule}
\end{figure}

Following~\cite{Aranson2006}, the time-evolution of the probability distribution function is assumed to be governed by a Boltzmann-like kinetic equation
\begin{align}
&\partial_{t}P(\vect{r},\phi) = D_{r}\partial_{\phi}^2P(\vect{r},\phi)+ \nabla_i D_{ij}\nabla_j P(\vect{r},\phi)\nonumber\\
&+\int d\vect{\xi}\int_{-\pi}^{\pi}d\omega \Biggl[
W\left(\vect{r}-\frac{\vect{\xi}}{2},\phi-\frac{\omega}{2};\vect{r}+\frac{\vect{\xi}}{2},\phi+\frac{\omega}{2}\right)  \nonumber \\
& \qquad\times P\left(\vect{r}-\frac{\vect{\xi}}{2},\phi-\frac{\omega}{2}\right)P\left(\vect{r}+\frac{\vect{\xi}}{2},\phi+\frac{\omega}{2}\right)\nonumber \\ 
& -W(\vect{r},\phi; \vect{r}-\vect{\xi},\phi-\omega) P \left(\vect{r},\phi\right) P \left(\vect{r}-\vect{\xi},\phi-\omega\right)\Biggr],
\label{MEfull} 
\end{align}
where $\nabla_i = \partial/\partial x_i$, $x_i$ are the Cartesian components of $\vect{r}$, and we use the Einstein summation convention; from now on we suppress the explicit time-dependence of $P$ for brevity. 
The first two terms in Eq.\eqref{MEfull} describe thermal rotational and translational diffusion of individual microtubule bundles, while the last two terms represent motor-mediated interactions between microtubule bundles. The first integral in Eq.\eqref{MEfull} is a gain term, which accounts for events where two microtubule bundles with different positions and orientations are re-oriented by the motors to assume position and orientation $\left(\vect{r},\phi\right)$. 
The specific form of this term encodes our assumptions about how motors and microtubules interact; the details of such interactions are summarised in Fig.~\ref{fig:Collision_rule}. Again following~\cite{Aranson2006}, we assume that after a re-orientation event both bundles align along the bisector of the original angle between them, while their centre of mass does not move in the process. 
To motivate the latter choice we note that a motor simultaneously attached to both bundles applies a pair of equal and opposite forces to the system -- i.e., it behaves as a force-dipole. Since the total force applied to the centre of mass is zero, its position is conserved. This assumption is in contrast with the work reported in~\cite{Liverpool2003,Liverpool2005,Ahmadi2006,Marchetti2013} where it was instead argued that directed motion of molecular motors along the microtubules can create a flow in the surrounding fluid that would result in microtubule self-propulsion, and, hence, the position of the centre of mass of two bundles can change during an interaction event. Such effects are rather difficult to quantify in dense suspensions of microtubules that are confined close to a boundary, as is typically the case in experiments, hence we neglect them here. The second integral in Eq.\eqref{MEfull} is a loss term, describing the process by which a bundle with the position and orientation $\left(\vect{r},\phi\right)$ leaves that configuration due to an interaction event with another bundle. The rate of both motor-induced processes is given by the function $W$ discussed below.

It is finally important to underscore that the pair-wise nature of the interaction terms in Eq.\eqref{MEfull} is {\it not} related to a dilute-limit assumption, as is often the case in Boltzmann-like kinetic theories, but rather stems from the fact that a molecular motor can only simultaneously attach to two microtubules~\cite{howard2001mechanics,Guerin2010}.
\\ 
\subsubsection{Long-wavelength expansion}
To proceed, we observe that without loss of generality the probability distribution function can be expanded in Fourier harmonics
\begin{align}
P\left(\vect{r},\phi\right) = \sum_{n=-\infty}^{\infty} P_n(\vect{r}) e^{i n \phi},
\label{Fharmonics}
\end{align}
where $P_{-n}^{*}(\vect{r})=P_n(\vect{r})$, since $P$ is real, and '$*$' denotes complex conjugation.
Next, we note that motor-mediated interactions between microtubules are short-ranged, and the integrand in Eq.\eqref{MEfull} is non-zero only when $|\vect{\xi}|\lesssim L$, independent of the particular form of the interaction strength $W$. Since we are interested in patterns that evolve slowly on scales comparable to $L$, we perform a gradient expansion of $P$ and keep terms up to fourth order. Projecting the resulting equation on the $s$-th Fourier harmonic yields the following equation,
\begin{widetext}
\begin{align}
& \partial_{t}P_s(\vect{r}) = -s^2 D_r P_s(\vect{r}) + \overline{\nabla_i D_{ij}\nabla_j P(\vect{r},\phi)}^{\, s} \nonumber \\
&+ \sum_{n,m=-\infty}^{\infty}\biggl[ \overline{I^{(0)}_{nm}}^{\,s} P_{n}P_{m} 
+ \frac{1}{2} \overline{I^{(1)}_{i,nm}}^{\,s} A_{i,nm}
+ \frac{1}{8} \overline{I^{(2)}_{ij,nm}}^{\,s} A_{ij,nm}
+ \frac{1}{48} \overline{I^{(3)}_{ijk,nm}}^{\,s} A_{ijk,nm}
+ \frac{1}{384} \overline{I^{(4)}_{ijkl,nm}}^{\,s} A_{ijkl,nm} + \cdots \biggr]  
\label{MEexpanded} \\
&- \sum_{n,m=-\infty}^{\infty} P_n \biggl[ \overline{J^{(0)}_{nm}}^{\,s} P_{m} 
- \overline{J^{(1)}_{i,nm}}^{\,s} \nabla_i P_m
+ \frac{1}{2} \overline{J^{(2)}_{ij,nm}}^{\,s} \nabla_i\nabla_j P_m
- \frac{1}{6} \overline{J^{(3)}_{ijk,nm}}^{\,s} \nabla_i\nabla_j\nabla_k P_m \nonumber \\
& \qquad\qquad \qquad + \frac{1}{24} \overline{J^{(4)}_{ijkl,nm}}^{\,s} \nabla_i\nabla_j\nabla_k\nabla_l P_m + \cdots \biggr], \nonumber 
\end{align}
\end{widetext}
where 
\begin{align}
\overline{\left( \dots \right)}^{\,s} = \frac{1}{2\pi}\int_{0}^{2\pi} e^{-i s \phi} \left( \dots \right),
\end{align}
and
\begin{align}
A_{i,nm} = P_n \nabla_i P_m - P_m \nabla_i P_n, \nonumber 
\end{align}
\begin{align}
& A_{ij,nm} = P_n \nabla_i \nabla_j P_m - 2 \left(\nabla_i P_n \right)\left(\nabla_j P_m \right) \nonumber \\
& \qquad + P_m \nabla_i \nabla_j P_n, \nonumber
\end{align}
\begin{align}
& A_{ijk,nm} = P_n \nabla_i \nabla_j\nabla_k P_m - 3 \left(\nabla_i P_n \right) \left(\nabla_j \nabla_k P_m \right) \nonumber \\
& \qquad + 3 \left(\nabla_i \nabla_j P_n \right) \left(\nabla_k P_m \right) - P_m \nabla_i \nabla_j\nabla_k P_n, \nonumber
\end{align}
\begin{align}
& A_{ijkl,nm} = P_n \nabla_i \nabla_j\nabla_k \nabla_l P_m - 4 \left(\nabla_i P_n \right) \left(\nabla_j \nabla_k \nabla_l P_m \right) \nonumber \\
&  \qquad + 6 \left(\nabla_i \nabla_j P_n \right)\left(\nabla_k \nabla_l P_m \right) - 4 \left(\nabla_i \nabla_j \nabla_k P_n \right) \left(\nabla_l P_m \right) \nonumber \\
&  \qquad + P_m \nabla_i \nabla_j\nabla_k \nabla_l P_n. \nonumber
\end{align}
In Eq.\eqref{MEexpanded} all $P_n$'s and $P_m$'s are functions of $\vect{r}$ and $t$. The interaction integrals are given by 
\begin{align}
& I^{(0)}_{nm} = e^{i(n+m)\phi}\int d\vect{\xi}\int_{-\pi}^{\pi}d\omega\,W_1 e^{i(m-n)\frac{\omega}{2}}, \nonumber \\
& I^{(1)}_{i,nm} = e^{i(n+m)\phi}\int d\vect{\xi}\int_{-\pi}^{\pi}d\omega\,W_1 e^{i(m-n)\frac{\omega}{2}}
\xi_i, \nonumber \\
& I^{(2)}_{ij,nm} = e^{i(n+m)\phi}\int d\vect{\xi}\int_{-\pi}^{\pi}d\omega\,W_1 e^{i(m-n)\frac{\omega}{2}}
\xi_i\xi_j,  \label{Iint}\\
& I^{(3)}_{ijk,nm} = e^{i(n+m)\phi}\int d\vect{\xi}\int_{-\pi}^{\pi}d\omega\,W_1 e^{i(m-n)\frac{\omega}{2}}
\xi_i\xi_j\xi_k,  \nonumber \\
& I^{(4)}_{ijkl,nm} = e^{i(n+m)\phi}\int d\vect{\xi}\int_{-\pi}^{\pi}d\omega\,W_1 e^{i(m-n)\frac{\omega}{2}}
\xi_i\xi_j \xi_k\xi_l, \nonumber
\end{align}
and
\begin{align}
& J^{(0)}_{nm} = e^{i(n+m)\phi}\int d\vect{\xi}\int_{-\pi}^{\pi}d\omega\,W_2 e^{-i m \omega}, \nonumber \\
& J^{(1)}_{i,nm} = e^{i(n+m)\phi}\int d\vect{\xi}\int_{-\pi}^{\pi}d\omega\,W_2 e^{-i m \omega} 
\xi_i, \nonumber \\
& J^{(2)}_{ij,nm} = e^{i(n+m)\phi}\int d\vect{\xi}\int_{-\pi}^{\pi}d\omega\,W_2 e^{-i m \omega} 
\xi_i\xi_j,  \label{Jint} \\
& J^{(3)}_{ijk,nm} = e^{i(n+m)\phi}\int d\vect{\xi}\int_{-\pi}^{\pi}d\omega\,W_2 e^{-i m \omega} 
\xi_i\xi_j\xi_k,  \nonumber \\
& J^{(4)}_{ijkl,nm} = e^{i(n+m)\phi}\int d\vect{\xi}\int_{-\pi}^{\pi}d\omega\,W_2 e^{-i m \omega} 
\xi_i\xi_j \xi_k\xi_l, \nonumber
\end{align}
where $\xi_i$ are the Cartesian components of $\vect{\xi}$, and we introduced
\begin{align}
& W_1\equiv W\left(\vect{r}-\frac{\vect{\xi}}{2},\phi-\frac{\omega}{2};\vect{r}+\frac{\vect{\xi}}{2},\phi+\frac{\omega}{2}\right), \nonumber \\
& W_2 \equiv W(\vect{r},\phi; \vect{r}-\vect{\xi},\phi-\omega).\nonumber
\end{align}
Eq.\eqref{MEexpanded} comprises an infinite hierarchy of equations for the Fourier harmonics $P_n(\vect{r},t)$. Its practical application relies on a strategy to reduce the number of relevant fields to just a few harmonics, and on the ability to calculate the interaction integrals for a particular function $W$. Our approach to both these issues is discussed below in Sections~\ref{closure1} and \ref{Q-t-c}, and in Section~\ref{kernel}), respectively.

\subsubsection{Diffusion terms}

The diffusion coefficients in Eq.\eqref{MEfull} are approximated by their values for a single rod of length $L$ and diameter $d$ moving in an infinite, three-dimensional fluid with viscosity $\eta$ \cite{DoiEdwards}
\begin{align}
D_r =12 \frac{k_B T }{\pi \eta L^3} \ln(L/d),
\label{Drot}
\end{align}
and
\begin{align}
D_{ij} = D_{\parallel} n_i(\phi) n_j(\phi) + D_{\perp} \left[ \delta_{ij} - n_i(\phi) n_j(\phi) \right],
\label{Drod}
\end{align}
where
\begin{align}
& D_{\perp} =\frac{k_B T }{4 \pi \eta L} \ln(L/d), \nonumber \\ 
& D_{\parallel} = 2D_{\perp}. \nonumber
\end{align}
Here, $T$ is the temperature of the solution, and $k_B$ is the Boltzmann constant.
Note that $D_r$ is four times larger than the value given by Doi and Edwards \cite{DoiEdwards} due to the difference in our choice of the angular variable (i.e. $\phi$ rather than $\mathbf{n}$). Using Eq.\eqref{Drod} in Eq.\eqref{MEexpanded}, and projecting onto the $s$-th Fourier harmonics, leads to the following contributions to the equations of motion
\begin{align}
& \qquad \partial_{t}P_s(\vect{r}) = -s^2 D_r P_s(\vect{r}) + \frac{D_{\parallel}+D_{\perp}}{2}\nabla^2 P_s(\vect{r}) \nonumber \\
& + \frac{D_{\parallel}-D_{\perp}}{4} \left( \nabla_x^2 - 2i\nabla_x\nabla_y - \nabla_y^2 \right) P_{s-2}(\vect{r}) \nonumber \\
& + \frac{D_{\parallel}-D_{\perp}}{4} \left( \nabla_x^2 + 2i\nabla_x\nabla_y - \nabla_y^2\right) P_{s+2}(\vect{r}) + \cdots,
\label{diffterms}
\end{align}
where '$\cdots$' denotes contributions from the interaction integrals discussed below.

Formally, Eqs.\eqref{Drot} and \eqref{Drod} limit the scope of Eq.\eqref{MEexpanded} to rather dilute suspensions far away from liquid-solid or liquid-liquid boundaries, while both assumptions are routinely violated in experiments \cite{Sanchez2012,Keber2014,Decamp2015,Needleman2017}. As we will see below, the kinetic theory equations that we are going to derive will only depend on the ratios $D_{\parallel}/D_r$ and  $D_{\perp}/D_r$ that are less sensitive to the local density of other microtubules and proximity of a boundary. We note, however, that a proper study of this effect is outside the scope of this work.

\subsubsection{Interaction kernel $W$}\label{kernel}

Within our kinetic theory, molecular details of motor-microtubules interactions are encoded at a coarse-grained level in the interaction function $W$. Its physical interpretation is given by Eq.\eqref{MEfull}, which identifies $W\left(\vect{r}_1,\phi_1;\vect{r}_2,\phi_2\right)$ as a rate at which two microtubular bundles at $\left(\vect{r}_1,\phi_1\right)$ and $\left(\vect{r}_2,\phi_2\right)$ are displaced and re-oriented by molecular motors. These changes in the bundles positions and orientations {\red occur when a molecular motor is attached to both bundles and moves along them. Therefore,} a motor-induced re-orientation event can only take place when the shortest distance between the bundles is not larger than the size of the motors. Since the latter is significantly smaller than the length of individual microtubules, or the typical size of the patterns formed by the suspension, see e.g.~\cite{howard2001mechanics,Sanchez2012}, we consider motors to be point-like. Under this assumption, $W$ is non-zero only when the bundles intersect in their original configuration. 
In turn, this implies that in real systems one of the bundles leaves the $xy$-plane of the suspension and deviates slightly into the third dimension. Such deviations are small compared either to $L$ or the typical pattern size, hence we will treat such bundles as intersecting in $2$D.
The {\red intersection condition} can be written as
\begin{eqnarray}
\mathbf{r}_1 + \mathbf{n}_1\frac{L}{2}\tau_1 = \mathbf{r}_2 + \mathbf{n}_2\frac{L}{2}\tau_2,
\label{param_intersection_point} 
\end{eqnarray}
where the left- and right-hand side of this equation is the position of the intersection point written with respect to the centre of mass (middle point) of either the first or the second bundle, i.e. $\mathbf{r}_1$ or $\mathbf{r}_2$. The microtubule orientation is given by $\mathbf{n}_i = \left( \cos\phi_i,\sin\phi_i\right)$, $i=1,2$. Here we have introduced the dimensionless contour lenghts $\tau_{1,2}$ that parametrise the position along each microtubule: $\tau=-1$ corresponds to the minus-end, and $\tau=1$ to the plus-end of the microtubule. By taking the cross product of Eq.\eqref{param_intersection_point} with either $\mathbf{n}_1$ or $\mathbf{n}_2$, the contour-length parameters can be found to be
\begin{align}
\label{tau1}
& \tau_1 = \frac{2}{L}\frac{\bigl(\left(\mathbf{r}_2-\mathbf{r}_1\right)\times\mathbf{n}_2\bigr)\cdot \mathbf{e}_z}
{\left(\mathbf{n}_1\times\mathbf{n}_2\right)\cdot \mathbf{e}_z}, \\
\label{tau2}
&\tau_2 =\frac{2}{L}\frac{\bigl(\left(\mathbf{r}_2-\mathbf{r}_1\right)\times\mathbf{n}_1\bigr)\cdot \mathbf{e}_z}{\left(\mathbf{n}_1\times\mathbf{n}_2\right)\cdot \mathbf{e}_z},
\end{align}
where $\mathbf{e}_z$ is a unit vector perpendicular to the $xy$-plane. Since $\left|\tau_{1,2}\right|$ should be smaller than unity, the intersection condition can equivalently be written as $\Theta\!\left(1 -\left|\tau_1\right|\right)\Theta\!\left(1 -\left|\tau_2\right|\right)\ne0$, where $\Theta$ is the Heaviside step function. 

Having established the condition {\red for bundle intersection}, we turn to modelling their re-orientation rate. Following Aranson and Tsimring \cite{Aranson2006}, we take this rate to be proportional to the local motor density at the intersection point. In the following we assume that the motors are abundant in the solution, and their dynamics of association/dissociation with the microtubules are much faster than the typical pattern-formation time. This was the case in several \emph{in-vitro} experiments (see \cite{Surrey2001}, for example). 

\begin{figure}[t]
\includegraphics[width=0.7\columnwidth]{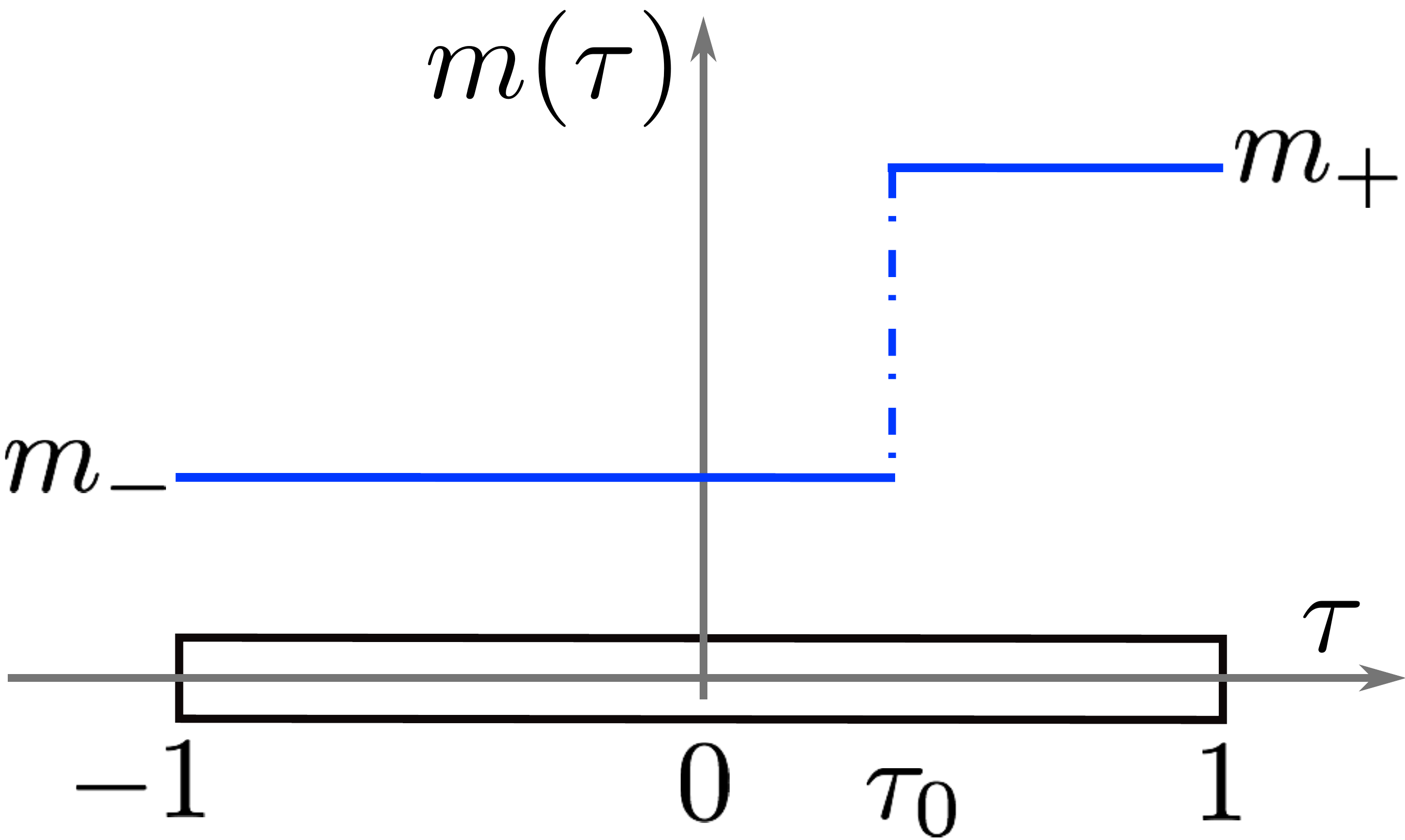}
\caption{Model anisotropic distribution of the molecular motors along a microtubular filament. }
\label{fig:Motor_step_function}
\end{figure}

With these assumptions, the motor distribution along individual microtubules instantaneously reaches its equilibrium profile. For plus-directed motors and under similar conditions, the equilibrium motor distribution was measured experimentally~\cite{Nishinari2005,Leduc2012}, and it was shown that the motor density stays low and approximately constant in the vicinity of the minus-end of the microtubules, until it rises sharply and saturates at another constant value close to the plus-end. This behaviour is corroborated by 1D non-equilibrium models~\cite{Parmeggiani2003,Nishinari2005,Aranson2006,Leduc2012} that relate this distribution to the formation of traffic jams at the plus-end. 
The equilibrium motor distribution $m(\tau)$, which gives the motor density at the contour length position $\tau$, can therefore be approximated by 
\begin{align}
m(\tau)= m_{-}+(m_{+}-m_{-})\Theta(\tau-\tau_0),
\end{align}
where $m_{-}$ and $m_{+}$ are the motor densities at the minus- and plus-ends, correspondingly, and $\tau_0$ sets the position of the transition between those values; see Fig.\ref{fig:Motor_step_function} for details.

The re-orientation rate can finally be written as
\begin{align}
&W\left(\vect{r}_1,\phi_1;\vect{r}_2,\phi_2\right) = G\,\Theta\!\left(1 -\left|\tau_1\right|\right)\Theta\!\left(1 -\left|\tau_2\right|\right) \nonumber \\
& \qquad\qquad \times \biggl\{ 1+\Xi \bigl[\Theta( \tau_1 - \tau_0)+\Theta(\tau_2 - \tau_0)\bigr]\biggr\},
\label{exactkernel}
\end{align}
where $\Xi=(m_{+}-m_{-})/(2 m_{-})$, and $\tau_{1,2}$ are given by Eqs.\eqref{tau1} and \eqref{tau2}. 
The constant $G$ is proportional to the motor properties, such as its processivity along the microtubules~\cite{howard2001mechanics,Cross2014}, and varies with the motor type. However, as we will demonstrate below, {\red $G$ can} be removed from the model by a rescaling of the dynamical fields.  While its value would be important to map the parameter values used in the equations of motion back to dimensional units, it plays no role in determining the phase diagram of our model. 
{\red Indeed,} the interaction function $W$ depends on two dimensionless parameters, $\tau_0$ and $\Xi$, where the latter quantifies the mismatch between the motor densities at the two ends of a microtubule. While it would be tempting to ignore this complexity {\red and set $\Xi=0$ for simplicity, previous work suggests this to be} a crucial ingredient of the theory. As was shown by Aranson and Tsimring \cite{Aranson2006} for their model, there is no interesting pattern formation taking place in the absence of the motor density mismatch, and only a trivial instability is present in that case (see below). A similar conclusion was reached by Marchetti, Liverpool and co-workers~\cite{Liverpool2003,Liverpool2005,Ahmadi2006,Marchetti2013}, where the analogous parameter was the motor speed anisotropy along a microtubule. We, therefore, consider $\Xi \ne 0$ below.

\section{Approximations in the Aranson-Tsimring theory}
\label{sec:AT}

In this section we review the approximations used by Aranson and Tsimring in~\cite{Aranson2006} to evaluate the integrals in Eqs.\eqref{Iint} and \eqref{Jint}, and to terminate the infinite hierarchy of coupled equations in Eq.\eqref{MEexpanded}. Here, we sketch their argument in some detail as it will be important in the further discussion. 

The first step involves replacing the exact interaction kernel $W$ in Eq.\eqref{exactkernel} with an  effective simplified kernel, which is given by
\begin{align}
&W_{AT}\left(\vect{r}_1,\phi_1;\vect{r}_2,\phi_2\right) = \frac{\tilde{G}}{b^2\pi}\exp\left(-\frac{\left(\mathbf{r}_1-\mathbf{r}_2\right)^2}{b^2}\right) \nonumber \\
& \qquad\qquad \times \left[ 1 - \frac{\beta}{L} \left(\mathbf{r}_1-\mathbf{r}_2\right)\cdot\left(\vect{n}_1-\vect{n}_2\right) \right],
\label{ATkernel}
\end{align}
where $b\sim L$ is a lengthscale, and $\tilde{G}$ is a motor-related constant, similar to $G$ in Eq.\eqref{exactkernel}. This expression replaces the complicated spatial and angular dependence of Eq.\eqref{exactkernel} with a Gaussian cut-off that, essentially, allows any interactions as long as the bundle centres of mass are separated by a typical distance set by $b$; the term in the brackets can be seen as the first terms of the Fourier expansion of the true angular dependence in Eq.\eqref{exactkernel}. The parameter $\beta$ is a measure of how anisotropic the motor distribution is along individual microtubules, and is analogous to $\Xi$ {\red in} Eq.\eqref{exactkernel}. The obvious benefit of this approximation is that the integrals in Eqs.\eqref{Iint} and \eqref{Jint} can now be evaluated analytically. In~\cite{Aranson2006} it is claimed that while not exact, Eq.\eqref{ATkernel} retains the main features of Eq.\eqref{exactkernel}. We demonstrate in the next Sections that together with the choice of the parameter $b$ made in \cite{Aranson2006}, {\red the approximation in Eq.\eqref{ATkernel}}  
{\red leads to a different phase diagram with respect to that obtained when the original kernel Eq.\eqref{exactkernel} is retained.}

The second approximation {\red developed in~\cite{Aranson2006} concerns the way to terminate the infinite hierarchy in Eq.\eqref{MEexpanded}.} 
To illustrate {\red this} strategy, we neglect spatial variations of the probability distribution and keep only its angular dependence. This approximation implies that the dominant mechanism of the instability in this system {\red should be} the appearance of orientational order, while the density fluctuations {\red are assumed} to be subdominant. {\rm The validity of this approximation will} be re-assessed after the same methodology is applied to the full system of equations with both the spatial and angular dependencies included. Using Eq.\eqref{ATkernel} in Eq.\eqref{MEexpanded}, and setting $\beta$ and the spatial gradients to zero, we obtain
\begin{align}
&\partial_{t}P_s = -s^2 D_r P_s  - 2\pi \tilde{G} P_0 P_s \nonumber \\
& \qquad \qquad+ \tilde{G} \sum_{m=-\infty}^{\infty} \frac{4\sin{\frac{\pi}{2}(2m-s)}}{2m-s} P_{s-m}P_m.
\label{only_angular}
\end{align}
Keeping only the first three Fourier harmonics in the expansion, this system of equations reads
\begin{align}
&\partial_{t}P_0 = 0, \\
& \partial_{t}P_1 = -D_r P_1  +\tilde{G} \left(8-2\pi\right) P_0 P_1 - \frac{8}{3} \tilde{G} P_{1}^{*}P_{2}, \\
& \label{P2}\partial_{t}P_2 = -4 D_r P_2  + 2\pi \tilde{G} \left(P_1^2 - P_0 P_2 \right),
\end{align}
where the first equation is the direct consequence of the total probability conservation. The isotropic solution of these equations is given by $P_1=P_2=0$, while the evolution of small perturbations around this state, $\delta p_1$ and $\delta p_2$, is governed by {\red the following equations},
\begin{align}
\partial_{t}\delta p_1 = \lambda_1 \delta p_1 , \qquad  \partial_{t}\delta p_2 = \lambda_2 \delta p_2,
\end{align}
where $\lambda_1 = -D_r  +\tilde{G} \left(8-2\pi\right) P_0$ and $\lambda_2=-4 D_r   - 2\pi \tilde{G} P_0$; here, $P_0$ is a constant. The isotropic solution becomes unstable with respect to perturbations $\delta p_1$ when $\lambda_1$ becomes positive, while perturbations in the second mode, $\delta p_2$, are decaying since $\lambda_2$ is always negative. Therefore, close to the instability threshold the dynamics of the second mode $P_2$ is {\red enslaved} to the dynamics of the linearly unstable field $P_1$ \cite{CrossHohenberg}, and $P_2$ can only acquire a non-zero value due to the non-linear forcing by the $P_1^2$ term in Eq.\eqref{P2}. Thus, $P_2$ quickly relaxes to the value set by the r.h.s. of Eq.\eqref{P2}
\begin{align}
P_2 = \frac{2\pi \tilde{G}}{4 D_r + 2\pi \tilde{G}  P_0} P_1^2.
\label{adiabaticp2}
\end{align}
As can be shown from Eq.\eqref{only_angular}, the same holds for all higher modes $P_m$ with $m>1$, where $P_m\sim O\left(P_1^m \right)$. Since close to the instability threshold the saturated value of $P_1$ is small, all higher harmonics are significantly smaller, and can be neglected. Therefore, {\red the authors of Ref.~\cite{Aranson2006} restricted} the infinite hierarchy Eq.\eqref{MEexpanded} to contain only the first three modes, $P_0$, $P_1$ and $P_2$, where the latter does not possess its own dynamics but is assumed to be well-approximated by the adiabatically-adjusted value given in Eq.\eqref{adiabaticp2}, even in the presence of spatial variations and non-zero $\beta$.

These approximations allow for Eq.\eqref{MEexpanded} to be converted into a system of partial differential equations for the hydrodynamic {\red (i.e., slowly varying)} fields $\rho(\vect{r})$ and $\vect{p}(\vect{r})$ defined by the moments of $P(\vect{r},\phi)$ {\red as follows,}
\begin{align}
\label{rho_def}
& \rho(\vect{r}) = \int_0^{2\pi} d\phi P(\vect{r},\phi) = 2\pi P_0(\vect{r}), \\
& \vect{p}(\vect{r}) = \frac{1}{2\pi}\int_0^{2\pi} d\phi\,\vect{n}(\phi) P(\vect{r},\phi) \nonumber\\ 
\label{p_def}
&\qquad = \left(\frac{P_{-1}(\vect{r})+P_1(\vect{r})}{2},\frac{P_{-1}(\vect{r})-P_1(\vect{r})}{2i} \right).
\end{align}
Here, $\rho(\vect{r})$ is the local density of microtubular bundles, and $\vect{p}(\vect{r})$ is proportional to their local orientation; note that $\vect{p}$ is not a unit vector.

To render equations dimensionless, time, space and the slow fields $\rho$ and $\vect{p}$ are scaled by $D_r^{-1}$, $L$ and $\tilde{G} L^2/D_r$, respectively. The final dimensionless equations used in~\cite{Aranson2006} read
\begin{align}
&\partial_t \rho=\nabla^2\left[\frac{\rho}{32}-\frac{B^2\rho^2}{16}\right]
-\frac{\pi B^2 H}{16} \bigg[ 3\nabla\cdot \left( \vect{p}\nabla^2\rho-\rho\nabla^2\vect{p} \right)\nonumber\\&
\qquad +2\partial_i\left(\partial_j\rho\partial_jp_i-\partial_i\rho\partial_jp_j \right) \bigg] 
-\frac{7\rho_0B^4}{256}\nabla^4\rho, 
\label{ATrhodot} \\ 
& \partial_t{\vect{p}} = \frac{5}{192}\nabla^2\vect{p}+\frac{1}{96}\nabla(\nabla\cdot\vect{p})+\left(\rho/\tilde\rho_{cr}-1\right)\vect{p}-\tilde A_0|\vect{p}|^2\vect{p}\nonumber\\
&\qquad -H\!\!\left[\frac{\nabla\rho^2}{16\pi}\!-\!\left(\pi-\frac{8}{3}\right)\!\vect{p}(\nabla\!\!\cdot\!\vect{p})-\!\frac{8}{3}(\vect{p}\cdot\nabla)\vect{p}\!\right] \nonumber \\
& \qquad +\!\frac{B^2\rho_0}{4\pi}\nabla^2\vect{p},
\label{ATpdot}
\end{align}
where $\rho_0$ is the conserved average density, $B=b/L$, $H=\beta B^2$, and $\tilde\rho_{cr} = \pi/(4 - \pi )$; the constant $\tilde A_0=16\pi/(3(\rho_0+4))$ and the corresponding term in Eq.\eqref{ATpdot} arise from the dimensionless version of Eq.\eqref{adiabaticp2}.

In addition to the approximations developed above, Eqs.\eqref{ATrhodot} and \eqref{ATpdot} {\red entail some additional} assumptions. First, the only non-linear terms (i.e. terms proportional to $H$) kept in these equations correspond to the lowest order non-zero terms in the gradient expansion (cubic and linear in gradients in the equations for $\rho$ and $\vect{p}$, respectively). This is done to ensure that both equations are coupled to each other. Additionally, Eq.\eqref{ATpdot} contains a series of terms quadratic in the gradient that are linearised around $\rho_0$, giving rise to the last term in that equation. {\red This linearisation is justified if there are only small density variations close to the instability threshold.} Finally, to ensure the absence of short-wavelength-instability, the fourth-order terms in the gradient expansion are again linearised around $\rho_0$ to yield the biharmonic term in Eq.\eqref{ATrhodot}.

{\red The analysis presented in the Aranson-Tsimring theory} suggests that Eqs.\eqref{ATrhodot} and \eqref{ATpdot} exhibit two linear instabilities: an isotropic-polar transition at $\rho_0=\tilde\rho_c$, where the system acquires a global polarisation $\vect{p}(\vect{r})=\text{const}$, while $\rho(\vect{r})=\rho_0$, and the bundling transition at $\rho_0=\tilde\rho_b\equiv1/\left(4B^2\right)$ with $\vect{p}(\vect{r})=0$, where the linearised diffusion-like term in Eq.\eqref{ATrhodot} becomes negative indicating the tendency of the system to accumulate disordered microtubular bundles in localised clusters; both instabilities are long-wavelength and set in at the scale of the system {\red size}.  By setting $B$ such that $\tilde\rho_c<\tilde\rho_b$, Aranson and Tsimring {red could show} numerically that for $\tilde\rho_c<\rho_0<\tilde\rho_b$ Eqs.\eqref{ATrhodot} and \eqref{ATpdot} exhibit a disordered quasi steady-state array of vortex and aster-like structures, dominated by vortices, at low $H$, and by asters, at larger $H$. For $\rho_0>\tilde\rho_b$, there is a competition between vortices, asters, and disordered high-density clusters at high values of $H$.

Below we systematically examine the assumptions {\red leading} to Eqs.\eqref{ATrhodot} and \eqref{ATpdot}. First, we devise a semi-analytical strategy to evaluate the integrals in Eq.\eqref{MEexpanded} with the exact interaction kernel Eq.\eqref{exactkernel} instead of the {\rm effective approximation} Eq.\eqref{ATkernel}. We will demonstrate that, {\red as a result, the bundling transition sets in} at lower density than the instability towards a globally ordered state, {\red substantially} changing the phase diagram. This can already be seen from comparing Eq.\eqref{ATkernel} with Eq.\eqref{exactkernel}: since $L$ is the only lengthscale that appears in the true interaction kernel, the parameter $b$ of the {\red approximate} kernel should only differ from $L$ by a factor of order unity, which implies $\tilde\rho_b\approx 1/4 < \rho_c$. Next we note that the terms that appear in Eqs.\eqref{ATrhodot} and \eqref{ATpdot} were selected on the basis of approximations {\red whose validity is difficult to control {\it a priori}:} as a result, close to the instability threshold the final equations combine terms {\red which effectively are of different orders.} 
We show how to systematically keep terms of the same order and that this requires {\red modification of} the closure given by Eq.\eqref{adiabaticp2}. Finally, we observe that in the absence of anisotropy in the interaction kernel, i.e. $H=0$, Eq.\eqref{ATrhodot} exhibits pathological behaviour for $\rho_0>\tilde\rho_b$, since there are no non-linear terms that can cut-off exponential growth of the linearly-unstable modes. The same problem persists at small values of $H$, while at large $H$ the non-linear coupling to the polarisation field limits the instability growth, as shown in~\cite{Aranson2006}. To cure this problem, which is more severe when $\tilde\rho_b<\tilde\rho_{cr}$, {\red here we explicitly account for} excluded volume interactions between the microtubular bundles that stabilise the dynamics even in the absence of the polarisation field.

\section{Excluded Volume Interactions} \label{Sec:ExcV}

In this Section, we incorporate the excluded volume interactions between microtubular bundles into the dynamic equation for the density. 
{\red To do so, we start} from the Smoluchowski equation, {\red similarly} to the work by Ahmadi \emph{et al.}~\cite{Ahmadi2006} and Baskaran and Marchetti~\cite{Baskaran2010}, and then incorporate these terms into the dynamical equations that we derived from the Boltzmann-like Eq.\eqref{MEfull}. Another approach is to introduce the excluded volume interactions directly in the Boltzmann-like equation~\cite{Bertin2015}, but this is more cumbersome. 
Formally, the two approaches are expected to be equivalent, but note their detailed comparison by Bertin \emph{et al.} \cite{Bertin2015}.

We begin by introducing the Onsager free energy~\cite{Onsager1949,Vroege1992,Mederos2014} for a collection of solid rods in terms of irreducible integrals~\cite{MayerMayer}
\begin{align}
&\frac{\mathcal{F}}{k_B T} = \int d\vect{r}\int d\phi \, P(\vect{r},\phi) \Biggl[ \ln{\Lambda^2 P(\vect{r},\phi)}- 1\Biggr] \nonumber \\
&-\frac{1}{2}\int d\vect{r}d\vect{r}' \int d\phi d\phi' P(\vect{r},\phi)P(\vect{r}',\phi') f(\vect{r},\phi;\vect{r}',\phi') \nonumber \\
&-\frac{1}{6}\int d\vect{r}d\vect{r}'d\vect{r}'' \int d\phi d\phi' d\phi'' P(\vect{r},\phi)P(\vect{r}',\phi') P(\vect{r}'',\phi'') \nonumber \\
& \times f(\vect{r},\phi;\vect{r}',\phi')f(\vect{r},\phi;\vect{r}'',\phi'')f(\vect{r}',\phi';\vect{r}'',\phi'') + \cdots.
\end{align} 
Here, $P$ is the probability distribution function, as in Section~\ref{sec:kineticmodel}, $\Lambda$ is the 
thermal de Broglie wavelength, and $f=\exp{\left(-U/k_B T\right)} -1$ is the Mayer function. The interaction potential $U$ between two microtubular bundles is infinite, when the bundles cross, and zero otherwise. In the absence of any external driving, the equilibrium probability distribution is given by \cite{DoiEdwards}
\begin{align}
\frac{\delta}{\delta P(\vect{r},\phi)} \Biggl[ \frac{\mathcal{F}}{k_B T}  - \lambda \int d\vect{r}\int d\phi \, P(\vect{r},\phi)  \Biggr] = 0,
\end{align}
where $\delta/\delta P(\vect{r},\phi)$ denotes a functional derivative w.r.t. $P(\vect{r},\phi)$, and we have introduced a Lagrange multiplier $\lambda$ to ensure that $P(\vect{r},\phi)$ satisfies the normalisation condition
\begin{align}
\int d\vect{r}\int d\phi \, P(\vect{r},\phi) = N.
\label{normalisation}
\end{align}
The solution to this equation can formally be written as 
\begin{align}
P(\vect{r},\phi)  = \text{const}\times \exp\left[-\frac{U_{sc}(\vect{r},\phi)}{k_B T} \right],
\end{align}
where the self-consistent potential $U_{sc}$ is a functional of the probability distribution function,
\begin{align}
&\frac{U_{sc}(\vect{r},\phi)}{k_B T}  = -\int d\vect{r}' \int d\phi' P(\vect{r}',\phi') f(\vect{r},\phi;\vect{r}',\phi') \nonumber \\
\label{Usc}
&-\frac{1}{2}\int d\vect{r}'d\vect{r}'' \int d\phi' d\phi'' P(\vect{r}',\phi') P(\vect{r}'',\phi'') \nonumber \\
& \qquad \times f(\vect{r},\phi;\vect{r}',\phi')f(\vect{r},\phi;\vect{r}'',\phi'')f(\vect{r}',\phi';\vect{r}'',\phi'') \\
& + \cdots, \nonumber
\end{align}
and the {\red constant, \text{const},}  is determined from the normalisation condition, Eq.\eqref{normalisation}. The two terms in Eq.\eqref{Usc} are the second and third irreducible integrals \cite{MayerMayer} that correspond to two-bundle and three-bundle interactions, respectively. As was shown by Ahmadi \emph{et al.} \cite{Ahmadi2006}, the first term leads to a contribution to the density equation proportional to $\nabla^2\rho^2$. When added to Eq.\eqref{ATrhodot}, for example, this contribution can limit the growth of the density fluctuations only for certain values of the parameter $B$, and in order to avoid this restriction we also include the three-bundle term in Eq.\eqref{Usc}, which, as we show below, leads to a contribution proportional to $\nabla^2\rho^3$ and provides a stabilisation mechanism for any density and any values of the parameters.

To evaluate the first integral in Eq.\eqref{Usc} we observe that the Mayer function $f(\vect{r},\phi;\vect{r}',\phi')$ is only non-zero when a bundle at $\vect{r}'$ with an orientation given by $\phi'$ intersects the test bundle at $\vect{r}$ with an orientation given by $\phi$, and in that case $f=-1$. The first integral, therefore, reduces to
\begin{align}
\int d\vect{r}' \int d\phi' P(\vect{r}',\phi'),
\end{align}
integrated over intersecting configurations only. To enumerate such configurations, we use the contour variables introduced in Eq.\eqref{param_intersection_point}, and write the condition of two bundles intersecting {\red on a plane} as
\begin{eqnarray}
\vect{r} + \vect{n}(\phi)\frac{L}{2}\tau = \vect{r}' + \vect{n}'(\phi')\frac{L}{2}\tau',
\label{intersectEV}
\end{eqnarray}
where $\tau$ and $\tau'$ are the dimensionless positions of the intersection point along the corresponding bundle; see discussion after Eq.\eqref{param_intersection_point} for details. When the bundles intersect, Eq.\eqref{intersectEV} can be used to change integration variables from $\vect{r}'$ to $\tau$ and $\tau'$, yielding
\begin{align}
& \frac{L^2}{4}\int_{-1}^{1}d\tau d\tau' \int_{0}^{2\pi} d\phi' P\left(\vect{r}+\frac{L}{2}\left( \vect{n}(\phi)\tau - \vect{n}'(\phi')\tau'\right),\phi'\right) \nonumber \\
& \qquad \times |\vect{e}_z \cdot \left( \vect{n}(\phi) \times \vect{n}'(\phi') \right)|,
\end{align}
where the last factor comes from the Jacobian of the transformation of variables. To proceed, we use the Fourier expansion of the probability density function, Eq.\eqref{Fharmonics}, in the integral above, and note, as before, that since we are interested in patterns evolving on spatial scales significantly larger than $L$, we can Taylor expand the Fourier modes $P_n\left(\vect{r}+\frac{L}{2}\left( \vect{n}(\phi)\tau - \vect{n}'(\phi')\tau'\right) \right)$ in gradients of $P_n(\vect{r})$. The leading contribution to this expansion comes simply from the zeroth order term $P_n(\vect{r})$ and we use this approximation in this calculation. Additionally, since we are interested in stabilising the dynamics of the density fluctuations, we will only keep $P_0$, ignoring all other Fourier harmonics, in the analysis below. The ignored contributions from higher Fourier modes and their spatial gradients have been discussed by Ahmadi \emph{et al.}~\cite{Ahmadi2006}. We will argue below that they are subdominant in the regime we are interested in. 

Proceeding with the approximations discussed above, the first integral in Eq.\eqref{Usc} becomes
\begin{align}
\frac{L^2}{4}\int_{-1}^{1}d\tau d\tau' \int_{0}^{2\pi} d\phi' \frac{\rho(\vect{r})}{2\pi} |\vect{e}_z \cdot \left( \vect{n}(\phi) \times \vect{n}'(\phi') \right)| \nonumber \\
 = L^2 \frac{\rho(\vect{r})}{2\pi} \int_{0}^{2\pi} d\phi'  |\sin(\phi-\phi')| = \frac{2}{\pi} L^2\rho(\vect{r}).
\end{align}

In a similar fashion, the second term in Eq.\eqref{Usc} can be written in terms of the dimensionless variables $\tau_{ij}$ that denote the position along the bundle $i$ of its crossing with the bundle $j$, where we have numbered the bundles with $(\vect{r},\phi)$, $(\vect{r}',\phi')$, and $(\vect{r}'',\phi'')$, as bundles $1$, $2$, and $3$, respectively. The conditions of simultaneous intersection of all three bundles is then
\begin{align}
&\vect{r} + \vect{n}(\phi)\frac{L}{2}\tau_{12} = \vect{r}' + \vect{n}'(\phi')\frac{L}{2}\tau_{21}, \nonumber \\
&\vect{r} + \vect{n}(\phi)\frac{L}{2}\tau_{13} = \vect{r}'' + \vect{n}''(\phi'')\frac{L}{2}\tau_{31}, 
\label{intersectEV3}\\
&\vect{r}' + \vect{n}'(\phi')\frac{L}{2}\tau_{23} = \vect{r}'' + \vect{n}''(\phi'')\frac{L}{2}\tau_{32}. \nonumber
\end{align}
As for the two-body interaction term, we use the first two conditions of Eq.\eqref{intersectEV3} to change variables from $\left(\vect{r}',\vect{r}''\right)$ to $\left(\tau_{12},\tau_{21},\tau_{13},\tau_{31} \right)$, and use the last condition to ensure that the bundles $2$ and $3$ cross. This yields for the second term in Eq.\eqref{Usc}
\begin{widetext}
\begin{align}
& \frac{1}{2}\left( \frac{\rho(\vect{r})}{2\pi}\right)^2 \left(\frac{L}{2}\right)^6 
\int_0^{2\pi} d\phi' d\phi'' |\sin(\phi-\phi')| |\sin(\phi-\phi'')|  |\sin(\phi'-\phi'')| \nonumber \\
& \qquad \times \int_{-1}^{1} d\tau_{12}d\tau_{21}d\tau_{13}d\tau_{31}d\tau_{23}d\tau_{32} 
 \,\delta\left(\frac{L}{2}\left\{ \vect{n}(\phi) \left(\tau_{12}-\tau_{13}\right) +  \vect{n}'(\phi') \left(\tau_{23}-\tau_{21}\right) +  \vect{n}''(\phi
') \left(\tau_{31}-\tau_{32}\right) \right\}\right) \nonumber \\
& = \frac{\rho(\vect{r})^2 L^4}{8 \pi^4} \int_0^{2\pi} d\phi' d\phi'' |\sin(\phi-\phi')| |\sin(\phi-\phi'')|  |\sin(\phi'-\phi'')| \nonumber \\
& \qquad \qquad \qquad  \times \int d\vect{k} \left(\frac{\sin(\vect{k}\cdot\vect{n}(\phi))}{\vect{k}\cdot\vect{n}(\phi)} \right)^2 \left(\frac{\sin(\vect{k}\cdot\vect{n}'(\phi'))}{\vect{k}\cdot\vect{n}'(\phi')} \right)^2 \left(\frac{\sin(\vect{k}\cdot\vect{n}''(\phi''))}{\vect{k}\cdot\vect{n}''(\phi'')} \right)^2,
\label{virial3}
\end{align}
\end{widetext}
where we replaced the two-dimensional Dirac delta-function by its integral representation \cite{lighthill1958book}
\begin{align}
\delta(\vect{a}) = \frac{1}{\left( 2\pi \right)^2} \int d\vect{k}\,e^{i\,\vect{k}\cdot\vect{a}},
\end{align}
and performed integration over $\tau$'s. We could not find a closed analytic form for the integral in Eq.\eqref{virial3}, and calculated it numerically by replacing the $[-\infty,\infty]\times[-\infty,\infty]$ integration range for $\vect{k}$ by $[-R,R]\times[-R,R]$, with $R$ appropriately large, but finite. We summed up the integrand on a a grid with $\Delta k_x=\Delta k_y =\Delta\phi=0.01$ and used $R=60$. Since the value of the integral should not depend on the absolute orientation of the first bundle, i.e. on the angle $\phi$, we also averaged the result over $\phi$ to increase accuracy. The resulting value is numerically very close to $2\pi^3$, and we will use that approximation below, for convenience. Finally, to third order in the bundle density the self-consistent potential becomes
\begin{align}
\frac{U_{sc}(\vect{r},\phi)}{k_B T} = \frac{2}{\pi} L^2\rho(\vect{r}) +  \frac{1}{4 \pi} L^4 \rho(\vect{r})^2.
\end{align}
The first term in the equation above was calculated by Ahmadi \emph{et al.}~\cite{Ahmadi2006}, and the three-dimensional version of the second term was discussed by Straley~\cite{Straley1973}.

Diffusion of a rod in an external potential is described by the Smoluchowski equation~\cite{DoiEdwards,Ahmadi2006}
\begin{align}
\frac{\partial P(\vect{r},\phi)}{\partial t} = \nabla_i D_{ij}\left( \nabla_j P(\vect{r},\phi) + P(\vect{r},\phi)  \nabla_j \frac{U}{k_B T}\right),
\end{align}
where $D_{ij}$ is given by Eq.\eqref{Drod}. Using $U_{sc}$ for the external potential, projecting onto the zeroth Fourier mode, and selecting only the terms containing the density, we arrive at the following contribution of the excluded volume effects to the dynamical equation for the density
\begin{align}
\partial_t \rho = \frac{D_{\parallel}+D_\perp}{2\pi}\nabla^2 \left( L^2 \rho(\vect{r})^2 + \frac{1}{6} L^4 \rho(\vect{r})^3\right) + \dots.
\label{EVtmp}
\end{align}
While the $\rho^3$ term provides stabilisation against the otherwise unbounded growth of the bundling instability, resolving the competition between the $\rho^2$ and $\rho^3$ terms numerically requires fine temporal resolutions, as at large timesteps the quadratic term can still lead to a finite-time blow-up due to an insufficient time for the qubic term to curb that growth. We, therefore, introduce a further approximation that allows us to avoid working with small timesteps by re-summing the virial expansion in Eq.\eqref{EVtmp} as
\begin{align}
& L^2 \rho(\vect{r})^2 + \frac{1}{6} L^4 \rho(\vect{r})^3 + \dots  \approx L^2 \rho(\vect{r})^2 e^{\frac{1}{6}L^2 \rho(\vect{r})},
\end{align}
where we added an infinite number of higher-order terms that mimic the effect of the higher order virial coefficients; their influence is small for sufficiently small densities, and their main function is to safe-guard against very fast growth of local density fluctuations in our numerical simulations presented below. Finally, the contribution of the excluded volume interactions to the density equation is written as
\begin{align}
\partial_t \rho = \frac{D_{\parallel}+D_\perp}{2\pi}\nabla^2  L^2 \rho(\vect{r})^2 e^{\frac{1}{6}L^2 \rho(\vect{r})} + \dots,
\label{EVfinal}
\end{align}
where $\dots$ denote the diffusion terms from Eq.\ref{diffterms}, and the terms originating from the interaction integrals are discussed next.

\section{Evaluation of interaction integrals}

In this Section we proceed by evaluating the interaction integrals from Eqs.\eqref{Iint} and \eqref{Jint} with the exact kernel Eq.\eqref{exactkernel}. As an example, we calculate the value of $I^{(1)}_{j,nm}$ which contains the same technical features shared by all other interaction integrals, whose values are given in Supplemental Material~\cite{SI}.

By introducing new variables $\chi = \psi-\phi$ and $\vect\zeta=\vect\xi/L$, the integral $I^{(1)}_{j,nm}$ can be written as
\begin{widetext}
\begin{align}
& \overline{I^{(1)}_{j,nm}}^s = G L^3 \frac{1}{2\pi} \int_0^{2\pi}d\phi \, e^{i(n+m)\phi}e^{-is\phi}\int_0^{2\pi}d\chi
	 \int_0^\infty d\zeta\,\zeta^2\int_{-\pi}^{\pi}d\omega W_1e^{i(m-n)\frac{\omega}{2}}
	 \left( \begin{matrix} \cos (\chi+\phi) \\ \sin ( \chi+\phi) \end{matrix} \right)_j,
\end{align}
where
\begin{align}
& W_1 = \Theta \left( |\sin\omega|-2\zeta\left|\sin\left(\chi-\frac{\omega}{2}\right)\right|\right)
\Theta\left(|\sin\omega|-2\zeta\left|\sin\left(\chi+\frac{\omega}{2}\right)\right|\right) \nonumber \\
&\qquad \times\bigg\{ 1+\Xi\left[ \Theta\left(-2\zeta\frac{\sin\left(\chi+\frac{\omega}{2}\right)}{\sin\omega}-\tau_0\right)+\Theta\left(-2\zeta\frac{\sin\left(\chi-\frac{\omega}{2}\right)}{\sin\omega}-\tau_0\right)\right]\bigg\}.
\label{IW1}
\end{align}
\end{widetext}
Projection onto the $s$-th Fourier harmonics yields
\begin{align}
	\frac{1}{2\pi}\int_{0}^{2\pi}e^{-i s \phi}e^{i\left(n+m\right)\phi}\left( \begin{matrix} \cos (\chi+\phi) \\ \sin ( \chi+\phi) \end{matrix} \right)_j d\phi=\nonumber\\
	\left(
	\begin{matrix}
	\left(e^{i\chi}\delta_{n,s-m-1}+e^{-i\chi}\delta_{n,s-m+1}\right)/2\\
	\left(e^{i\chi}\delta_{n,s-m-1}-e^{-i\chi}\delta_{n,s-m+1}\right)/(2i)
	\end{matrix}
	\right)_j,
\end{align}
and the spatial components of $I^{(1)}_{j,nm}$ can be expressed as
\begin{align}
    \overline{I^{(1)}_{x,nm}}^s&=L^3G\frac{B^{(1)}_{s,m}\delta_{n,s-m-1}+B^{(2)}_{s,m}\delta_{n,s-m+1}}{2},\nonumber\\
    \overline{I^{(1)}_{y,nm}}^s&=L^3G\frac{B^{(1)}_{s,m}\delta_{n,s-m-1}-B^{(2)}_{s,m}\delta_{n,s-m+1}}{2i},
\end{align}
where 
\begin{align}
    B^{(1)}_{s,m}=& G^{-1}\int_0^{2\pi}d\chi
	 \int_0^\infty d\zeta \zeta^2\int_{-\pi}^{\pi}d\omega W_1e^{i(2m-s+1)\frac{\omega}{2}}
	 e^{i\chi},\nonumber\\
	B^{(2)}_{s,m}=& G^{-1}\int_0^{2\pi}d\chi
	  \int_0^\infty d\zeta \zeta^2\int_{-\pi}^{\pi}d\omega W_1e^{i(2m-s-1)\frac{\omega}{2}}
	 e^{-i\chi},\nonumber
\end{align}
are functions of $\Xi$ and $\tau_0$. The structure of Eq.\eqref{IW1} suggests that each of these integrals can be split into two contributions
\begin{align}
     B^{(k)}_{m,s}(\Xi,\tau_0)=B^{(k)iso}_{m,s}+\Xi\,B^{(k)ani}_{m,s}(\tau_0),
\end{align}
where $B^{(k)iso}_{m,s}$ is a number associated with the isotropic (i.e. $\Xi$-independent) part of the kernel Eq.\eqref{exactkernel}, while $B^{(k)ani}_{m,s}$ is a function of $\tau_0$. We evaluate these contributions numerically, as explained below. 

\begin{figure}[t]
	\centering
	\includegraphics[width=3.4in,scale=1]{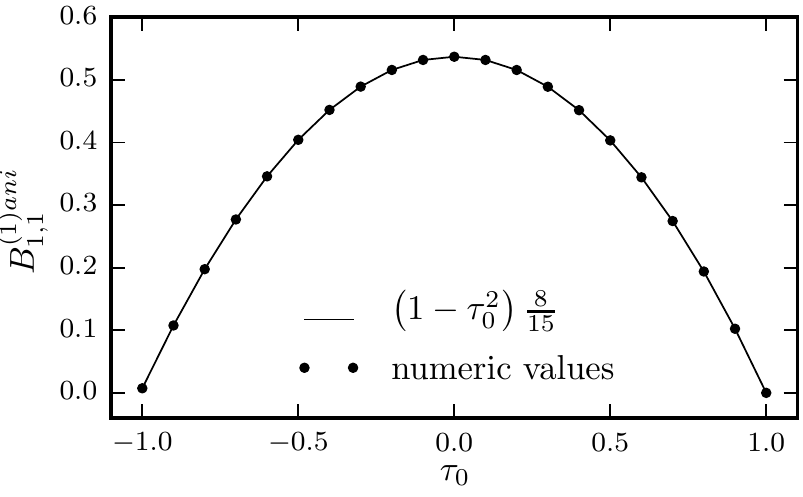}
	\caption{$B^{(1)ani}_{1,1}$ as a function of $\tau_0$.}
	\label{fig:B1ani}
\end{figure}

We illustrate our method by calculating $B^{(1)}_{1,1}$. In this case, numerical integration on a grid with $\Delta \zeta=\Delta \omega=\Delta\chi=0.01$ gives $B^{(1)iso}_{1,1}=0$. We note, however, that in general the isotropic contributions to this and other integrals are not necessarily zero for all values of the indices. To evaluate the anisotropic part $B^{(1)ani}_{1,1}(\tau_0)$, we perform a similar numerical integration for a range of $\tau_0$ from the interval $\left[-1,1\right]$, and plot the resulting values in Fig.\ref{fig:B1ani} (solid circles). We observe that these values are well-approximated by 
\begin{align}
B^{(1)ani}_{1,1}(\tau_0)=\frac{8}{15}\left(1-\tau_0^2\right),	
\end{align} 
as can be seen from Fig.\ref{fig:B1ani} (solid line). We therefore obtain
\begin{align}
    B^{(1)}_{1,1}(\Xi,\tau_0)=\frac{8}{15}\Xi\left(1-\tau_0^2\right).
\end{align}

All other integrals $I$'s and $J$'s, Eqs.\eqref{Iint} and \eqref{Jint}, are evaluated in the same way. For all these integrals, the anisotropic contributions are simple polynomials in $\tau_0$ that are readily {\red guessed}, while their prefactors and the isotropic contributions are well approximated by ratios of simple integers (see Supplemental Material~\cite{SI} for detail).

\section{Hydrodynamic equations}
\label{sec:models}

We now have all the ingredients to formulate our version of the equations of motion for the hydrodynamic fields. As mentioned above, our approach differs from the work {\red of~\cite{Aranson2006}} in several important ways, and we will show that this significantly changes the phase diagram of the system. In order to be able to attribute the changes observed to a particular aspect of our theory, we use the following approach. First, we use our values of the interaction integrals calculated with the exact kernel in Eq.\eqref{MEexpanded} combined with a closure strategy employed {\red in~\cite{Aranson2006}},  see Eq.\eqref{adiabaticp2}. Then we repeat the same derivation but with a different closure devised to keep only the terms that are relevant in the vicinity of the instability onset. In both cases we add the excluded volume terms to the equation for the density to be able to resolve the dynamics in the presence of a bundling instability, as discussed in Section~\ref{Sec:ExcV}.

\subsection{Aranson-Tsimring Closure}\label{closure1}
Here, we repeat the derivation from Section~\ref{sec:AT} with our values of the interaction integrals. The equations are rendered dimensionless by scaling time, space and the Fourier harmonics of $P$ by $D_r^{-1}$, $L$ and $G L^2/D_r$, respectively. In Eq.\eqref{MEexpanded}, we keep only the first Fourier harmonics, $P_0$, $P_{\pm 1}$, and $P_{\pm 2}$, but drop any gradient of $P_{\pm 2}$. For the second Fourier harmonics, Eq.\eqref{MEexpanded} is an algebraic equation that is solved by $P_{\pm2}=A_0P_{\pm1}^2$, similar to the closure Eq.\eqref{adiabaticp2}, while for the density and polarisation, Eqs.\eqref{rho_def} and \eqref{p_def}, we obtain
\begin{widetext}
\begin{align}
&\partial_t \rho =\nabla^2\left[\frac{\rho}{32}-\frac{(1+a_3)\rho^2}{48\pi}\right]+\frac{1}{32\pi}\alpha\nabla^2 \left( \rho^2 e^{\frac{\alpha\rho}{6}}\right)  -\frac{91}{69120\pi}\left(1+a_5\right)\rho_0\nabla^4\rho \nonumber \\
&\qquad\qquad -\frac{1}{240}a_4\left[
		3\nabla\cdot
		\left(
			\vect{p}\nabla^2\rho-\rho\nabla^2\vect{p}
		\right)
		+2\partial_i\left(
		\partial_j\rho\partial_jp_i-\partial_i\rho\partial_jp_j
		\right)
	\right],
\label{OurATrho} \\
	\partial_t \vect{p} =& -\vect p+\frac{5}{192}\nabla^2\vect{p}+\frac{1}{96}\nabla(\nabla\cdot\vect{p})
	+\left(1+a_1\right)\left(\frac{2}{3\pi}\rho\vect p-\frac{28}{15}A_0|\vect{p}|^2\vect{p}\right)\nonumber\\
	&-a_2\left[\frac{\nabla\rho^2}{32\pi^2}-\frac{1}{20}\vect{p}(\nabla\cdot\vect{p})-\frac{9}{20}(\vect{p}\cdot\nabla)\vect{p}+\frac{1}{24}\nabla\left(\vect p\cdot\vect p \right)\right]
	+\left(1+a_3\right)\frac{\rho_0}{40\pi}\bigg(\nabla^2\vect p +\frac{2}{9}\nabla(\nabla\cdot\vect p) \bigg).
\label{OurATp}
\end{align}   
\end{widetext}
Here,
\begin{equation}\label{relationP2P1our}
	A_0=\frac{3\pi}{3\pi\left(1+a_1\right)^{-1}+\rho_0},
\end{equation}
and
\begin{align}
	&a_1=\Xi\left(1-\tau_0\right),\nonumber\\
	&a_2=\Xi (1 - \tau_0^2),\nonumber\\
	&a_3=\Xi\left(1-\tau_0\left(1+\tau_0^2\right)/2\right),\\
	&a_4=\Xi\left(1-\tau_0^2\left(1+\tau_0^2\right)/2\right),\nonumber\\
	&a_5=\Xi\left(1-\tau_0\left(1+\tau_0^2\right)^2/2\right). \nonumber
\end{align}

In Eq.\eqref{OurATrho}, the term proportional to $\alpha$ is the dimensionless version of the excluded volume contribution from Eq.\eqref{EVfinal}, where $\alpha = D_r/G$. {\red This quantity} can be understood as a ratio of two timescales, $t_m/t_r$, where $t_m\sim G^{-1}$ is a typical time {\red over which a} bundle changes its orientation due to the activity of molecular motors, while $t_r\sim D_r^{-1}$ is a typical re-orientation time due to rotational diffusion. In the absence of motor activity, $\alpha$ becomes very large, and the excluded volume term prevents formation of any significant density fluctuations. In the motor-activity-dominated regime, $\alpha$ is small, and this regime is the focus of the rest of this work.

We also note that apart from the $\nabla\left(\vect p \cdot\vect p\right)$ term and the excluded volume contribution, Eqs.\eqref{OurATrho} and \eqref{OurATp} have the same tensorial structure as the equations {\red in the Aranson-Tsimring theory}, Eqs.\eqref{ATrhodot} and \eqref{ATpdot}. {\red Perhaps surprisingly, though}, the differences in their dependence on the parameters of the kernel, and different numerical prefactors are sufficient to produce a rather different phase diagram, as we discuss in Section \ref{sec:results}.

\subsection{Self-consistent closure and Q-tensor}
\label{Q-t-c}

The inherent problem of the previous closure is that it combines terms that are of various {\red orders (equivalently, degrees of smallness)} close to the instability threshold. Dropping spatial gradient in the equation for the second Fourier harmonics of $P$ implies that, to obtain a coupled system, the density equation should contain third-order spatial gradients, whilst only first-order terms are sufficient in the polarisation equation. To address this inconsistency, we employ a systematic procedure of deriving hydrodynamic equations that was originally developed for Ginzburg-Landau-like amplitude equations in pattern formation~\cite{CrossHohenberg} and was recently applied in the context of self-propelled rods~\cite{Peshkov2012,Peshkov2014} and microtubule-motor mixtures~\cite{Ziebert2005}.

Similar to Eq.\eqref{ATpdot}, Eq.\eqref{OurATp} suggests that a uniformly polarised state becomes stable above some $\rho_{cr}$, given by the time- and spatially-independent version of Eq.\eqref{OurATp}. If we introduce $\epsilon^2=\rho_0-\rho_{cr}$, balancing the terms in Eq.\eqref{OurATp} implies that  $|{\vect p}|\sim \epsilon$, $\nabla \sim \epsilon$, $\partial_t \sim \epsilon^2$, and the deviation of the density $\rho({\vect r},t)$ from its average value $\rho_0$ scales as $\delta \rho({\vect r},t) \equiv \rho({\vect r},t) -\rho_0\sim\epsilon^2$. Using these scalings we can see that the coupling terms, i.e. the terms proportional to $a_4$, in Eq.\eqref{OurATrho} contain a term proportional to $\rho_0 \nabla^2 \left( \nabla\cdot{\vect p}\right)\sim\epsilon^4$, while the rest of the coupling terms are $\sim \epsilon^6$.  Moreover, this scaling implies that ignoring spatial gradients of $P_2$ or spatial gradients in the equation for $P_2$ is not justified as, for example, the term $\nabla_i p_j$ is of the same order as $p_i p_j$, used in the algebraic closure above. Therefore, here we re-derive the equation for $P_2$ keeping all the terms that are $\sim \epsilon^2$. 

To simplify the notation, we introduce the \emph{so-called} $Q$-tensor that is proportional to the second Fourier harmonics of $P({\vect r},t)$,
\begin{equation}
Q_{ij}(\vect r)=\frac{1}{\pi}\int_{0}^{2\pi}\left(n_in_j-\frac{1}{2}\delta_{ij}\right) P(\vect r,\phi)d\phi.
\end{equation}
{\red The two independent components of the $Q$-tensor can be explicitly written as follows,}
\begin{align}\label{QtroughP2}
 Q_{xx}(\vect{r}) &= \frac{P_2(\vect r)+P_{-2}(\vect r)}{2},\nonumber\\
 Q_{xy}(\vect{r}) &= \frac{P_{-2}(\vect r)-P_2(\vect r)}{2i}.
\end{align}
Note that $Q_{yy} = -Q_{xx}$ {\red and $Q_{yx}=Q_{xy}$ as the $Q$-tensor is traceless and symmetric}. Keeping the terms proportional to $\epsilon^2$ in Eq.\eqref{MEexpanded} for the second harmonics, we obtain
\begin{align}
&Q_{ij}=\frac{1}{1+\frac{1}{3\pi}\rho_0(1+a_1)}\Bigg[(1+a_1)\Big\{ 2 p_ip_j - (\vect p\cdot \vect p)\delta_{ij}\Big\} \nonumber \\
&\qquad\qquad  +a_2\frac{\rho_0}{48\pi}\Big\{\partial_ip_j+\partial_jp_i-\delta_{ij}(\nabla\cdot\vect p)\Big\}\Bigg],
\label{Qijclosure}
\end{align}
which, in the absence of spatial gradients, is the same as the closure used above. Similarly, keeping the leading terms in $\epsilon$, which are proportional to $\epsilon^3$ and $\epsilon^4$ for the first and the zeroth harmonics, respectively, we arrive at the following dynamical equations
\begin{widetext}
\begin{align}
&\partial_t{\rho}=\nabla^2\left[\frac{\rho}{32}-\frac{(1+a_3)\rho^2}{48\pi}\right]+\frac{1}{32\pi}\alpha\nabla^2 \left( \rho^2 e^{\frac{\alpha\rho}{6}}\right)  -\frac{91}{69120\pi}\left(1+a_5\right)\rho_0\nabla^4\rho \nonumber \\
& \qquad\qquad + \left[ \frac{\pi}{48}  -\frac{1}{36}(1+a_3)\rho_0 \right]
	\partial_i\partial_j Q_{ij} +\frac{1}{80}a_4\rho_0 \nabla^2 \left(\nabla\cdot\vect{p}\right),
\label{ourRhowithQ} \\
&\partial_t p_i= - p_i+\frac{5}{192}\nabla^2 p_i+\frac{1}{96}\nabla_i(\nabla\cdot\vect{p})
	+\left(1+a_1\right)\left(\frac{2}{3\pi}\rho\,p_i -\frac{28}{15}Q_{ij}p_j\right)\nonumber\\
& \qquad\qquad -a_2\left[\frac{1}{16\pi^2}\rho_0\partial_i\rho -\frac{1}{20}p_i(\nabla\cdot\vect{p})-\frac{9}{20}(\vect{p}\cdot\nabla)p_i+\frac{1}{24}\nabla_i\left(\vect p\cdot\vect p \right)+\frac{1}{240}\rho_0\partial_kQ_{ik}\right] \nonumber \\
& \qquad\qquad +\frac{\left(1+a_3\right)\rho_0}{8}\bigg[\frac{1}{5\pi}\nabla^2 p_i+\frac{2}{45\pi}\nabla_i(\nabla\cdot\vect p)\bigg].
\label{ourPwithQ} 	
\end{align}
\end{widetext}
In Eq.\eqref{ourRhowithQ} we kept several terms that are inconsistent with this approximation scheme. While being of higher order than the rest of the equation, they represent the lowest order terms responsible for a particular effect. Thus, we keep the term that causes the bundling instability and the excluded volume term that saturates it, and we follow Aranson and Tsimring~\cite{Aranson2006} in keeping the bilaplacian term that selects the lengthscale of the bundling instability. This system of equations is the central result of our paper.

\section{Results}
\label{sec:results}

In this Section we present analysis of the dynamical behaviour exhibited by the models derived above. For convenience, we will be referring to Eqs.\eqref{OurATrho} and \eqref{OurATp} as the Aranson-Tsimring-closure (ATC) model, and to Eqs.\eqref{ourRhowithQ} and \eqref{ourPwithQ} -- as the Q-tensor-closure (QC) model. First, we perform a linear stability analysis of the homogeneous and isotropic base state for both models and determine the regions of the parameter space where non-trivial behaviour can be expected. Then we perform direct numerical simulations of the ATC and QC models in these parts of the parameter space and discuss the resulting patterns.

\subsection{Linear stability analysis}
We start by observing that both models support exact solutions in the form of a homogeneous state with $\rho(\vect{r},t)=\rho_0$ and $\vect{p}(\vect{r},t)=\vect{P}$, as was already mentioned above. In both cases, the evolution equation for $\vect{P}$ is given by
\begin{align}
\partial_t P_i = \left[-1 +\left(1+a_1\right)\left(\frac{2}{3\pi}\rho-\frac{28}{15}A_0 P^2\right) \right] P_i,
\label{Phomog}
\end{align}   
where $P = |\vect{P}|$. Trivially, ${\vect P}=0$ is always a solution to this equation for any density. For densities larger than 
\begin{align}
\rho_{cr} = \frac{3 \pi }{2 (1+a_1)},
\label{rhocr}
\end{align}
the isotropic solution loses its stability, as Eq.\eqref{Phomog} suggests, and another solution sets in with
\begin{align} 
P = \frac{1}{1+a_1}\sqrt{\frac{15}{28}}\sqrt{\left(\frac{\rho_0}{\rho_{cr}}-1\right)\left(\frac{\rho_0}{2\rho_{cr}}+1\right)},
\label{Pglob}
\end{align}
and a random orientation selected through a spontaneous symmetry breaking. We refer to this solution as the \emph{globally-ordered} state.

\begin{figure}[t]
\includegraphics[width=1.0\columnwidth]{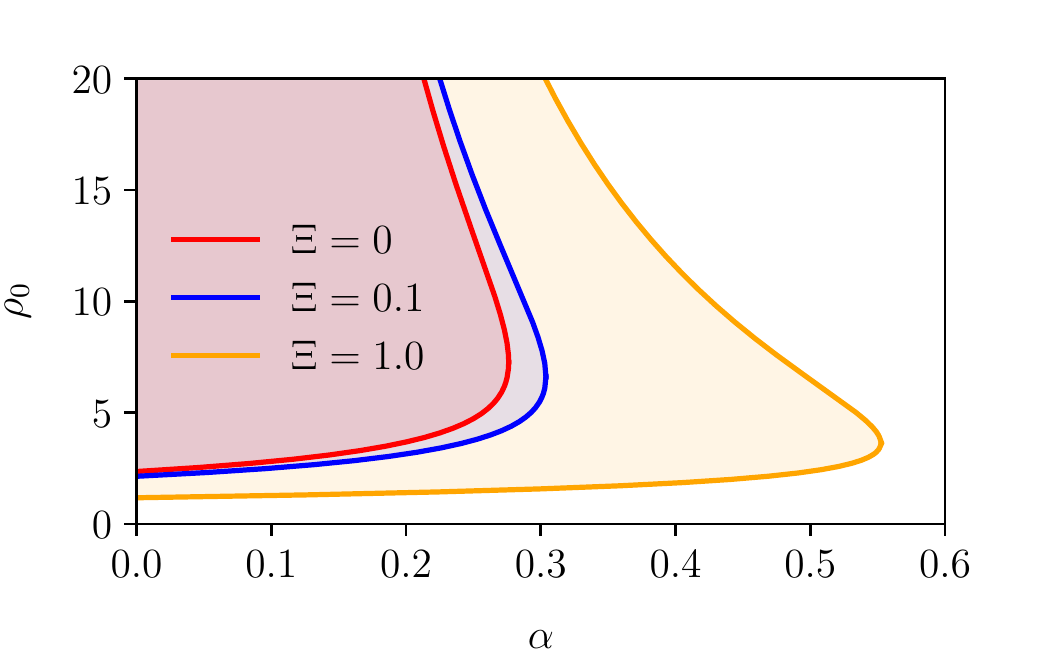}
\caption{Regions of existence of the bundling instability for $\tau_0=0$ and various values of $\Xi$. The solid lines are solutions to Eq.\eqref{rhob}, while the shaded regions indicate where the homogeneous and isotropic state is unstable with respect to density fluctuations (the bundling instability). {\red The solid lines can therefore be seen as spinodal lines, and the shaded areas as regions of phase separation.}}
\label{FigRhoAlpha}
\end{figure}

The homogeneous and isotropic state with $\rho(\vect{r},t)=\rho_0$ and $\vect{p}(\vect{r},t)=0$ is also unstable with respect to density fluctuations, as was already mentioned above; there, it was referred to as a \emph{bundling} instability. Assuming small spatial variations of the density profile $\rho(\vect{r},t)=\rho_0 + \delta\rho(t) e^{i(k_x x + k_y y)}$ and the absence of polarisation fluctuations, the linear dynamics of the density perturbations are given by $\partial_t{\delta\rho}=\lambda_{b}(k)\delta\rho$, where
\begin{align}
& \lambda_{b}(k)=\left[-\frac{1}{32}+\frac{(1+a_3)\rho_0}{24\pi} - \frac{\alpha\rho_0}{192\pi} e^{\frac{\alpha\rho_0}{6}} \left(12+\alpha\rho_0 \right)\right] k^2  \nonumber \\
& \qquad\qquad\qquad -\frac{91}{69120\pi}\left(1+a_5\right)\rho_0 k^4,
\label{eigen_b}
\end{align}
and $k^2=k_x^2+k_y^2$. For a selected wavevector, density perturbations grow when $\lambda_b(k)$ becomes positive, which can only happen when the coefficient of $k^2$ is positive, since the prefactor of $k^4$ is negative for realistic values of $\tau_0$. {\red Therefore this} instability sets in at a critical density $\rho_b$, given by
\begin{align}
-\frac{1}{32}+\frac{(1+a_3)\rho_b}{24\pi} - \frac{\alpha\rho_b}{192\pi} e^{\frac{\alpha\rho_b}{6}} \left(12+\alpha\rho_b \right) = 0,
\label{rhob}
\end{align}
which in the absence of the excluded volume, $\alpha=0$, becomes $\rho_{b} = 3 \pi/( 4(1+a_3))$, similar to the expression obtained {\red in~\cite{Aranson2006}}. 

In Fig.\ref{FigRhoAlpha}, we plot the solutions of Eq.\eqref{rhob} as a function of the excluded volume strength $\alpha$ for fixed values of the asymmetry parameter $\Xi$. For any value of $\Xi$, there exist two regions of this parameter space. For large values of $\alpha$ there is no bundling instability as strong excluded volume effects preclude growth of any density variations. {\red Instead,} for smaller values of $\alpha$ there is a band of density values (the shaded regions in Fig.\ref{FigRhoAlpha}), where the bundling instability exists. The upper boundary of this band goes to infinity when $\alpha$ approaches zero. 

\begin{figure}[!h]
\includegraphics[width=1.0\columnwidth]{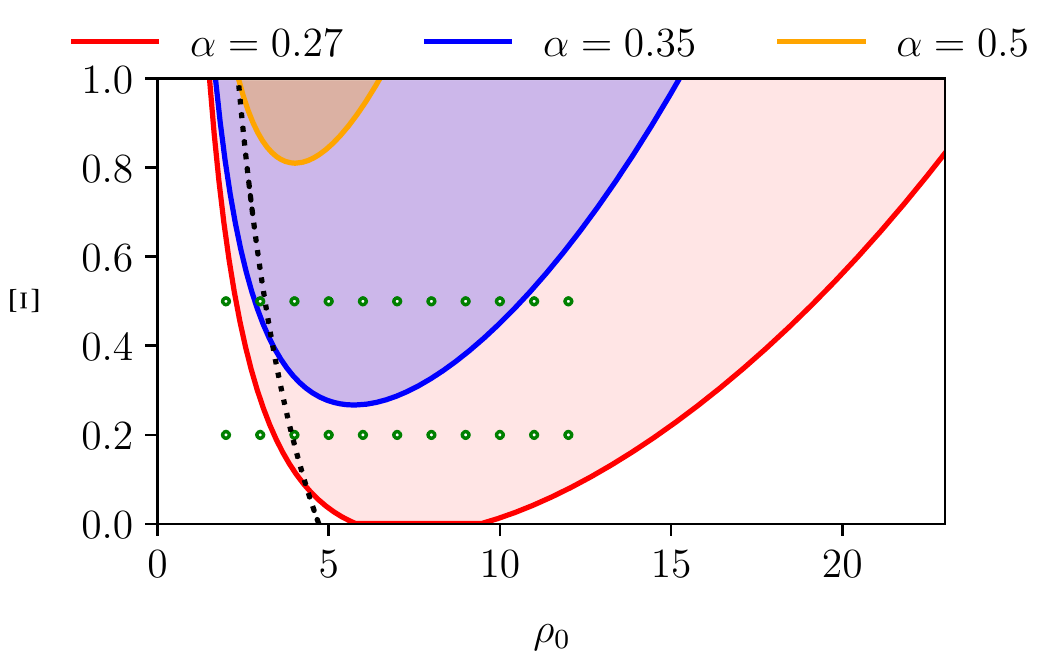}
\caption{The data from Fig.\ref{FigRhoAlpha} replotted as $\Xi$ vs. $\rho_0$ graph for $\tau_0=0$ and various values of $\alpha$. The dotted black line is the onset of global order, given by Eq.\eqref{rhocr}. Green circles indicate point for which we perform direct numerical simulations with the ATC and QC models: $\tau_0=0$, $\alpha=0.35$, $\Xi = 0.2$ and $\Xi=0.5$ with $\rho_0=2$, $3$, $\dots$, $12$. }
\label{FigRhoKsi}
\end{figure}

Since $\alpha$ sets the strength of the excluded volume interactions, which is just a geometric effect not related to motor activity, we fix {\red its value}, and treat $\Xi$ and $\rho_0$ as the control parameters. In Fig.\ref{FigRhoKsi}, we plot the instability boundaries found above in terms of these control parameters.
The dotted black line in Fig.\ref{FigRhoKsi} is the critical density $\rho_{cr}$, given by Eq.\eqref{rhocr}, while the solid lines, given by Eq.\eqref{rhob}, enclose the region of the bundling instability (shaded regions in Fig.\ref{FigRhoKsi}). As $\alpha$ increases, the bundling instability is pushed towards larger values of $\Xi$, but is always present. We, therefore, select a representative case of $\alpha=0.35$ (blue line and the blue shaded region in Fig.\ref{FigRhoKsi}),  and perform direct numerical simulations of the ATC and QC models for a range of densities and fixed motor asymmetry parameter $\Xi=0.2$ and $\Xi=0.5$. The former case exhibits only the instability towards a globally ordered state, while the latter case has both types of instability. The densities we use in our simulations are denoted by green circles in Fig.\ref{FigRhoKsi}.

Finally, we note that a full linear stability analysis (see below) shows that the transition to global order and the bundling instability are the only instabilities of the homogeneous and isotropic state for both models.

\begin{figure}[t]
\subfloat[$\rho_0=5$]{%
  \includegraphics[width=0.5\columnwidth]{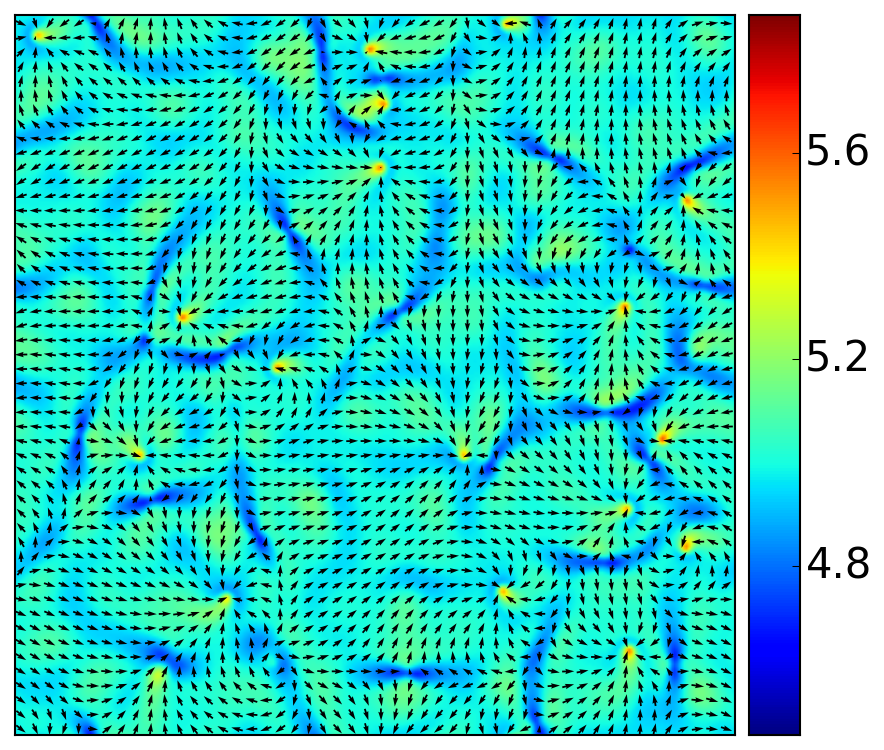}%
  \label{FigATC02r4}
}
\subfloat[$\rho_0=12$]{%
  \includegraphics[width=0.51\columnwidth]{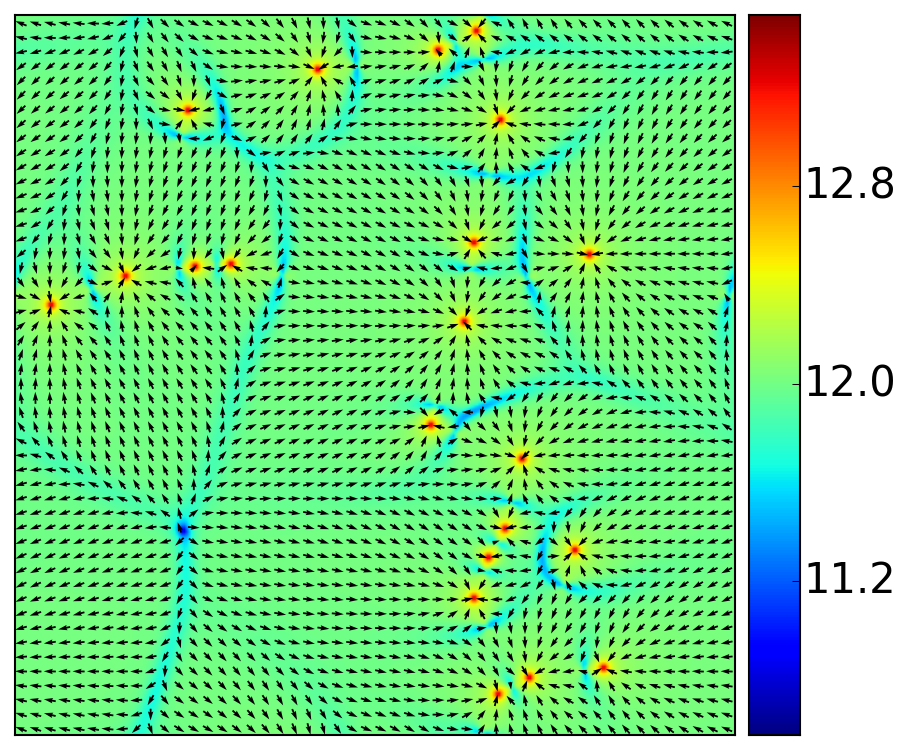}%
   \label{FigATC02r7}
}
\caption{Instantaneous snapshots from the direct numerical simulations of the ATC model with $\tau_0=0$, $\alpha=0.35$, and $\Xi = 0.2$. }
\label{FigATC02}
\end{figure}

\subsection{Direct numerical simulations}

To explore the nonlinear behaviour of the ATC and QC models, we perform direct numerical simulations of Eqs.\eqref{OurATrho} and \eqref{OurATp}, and of Eqs.\eqref{ourRhowithQ} and \eqref{ourPwithQ} in the parts of the parameter space identified above. We discretise spatial derivatives by second-order finite-differences, and employ a second-order predictor-corrector method for time integration~\cite{AbramowitzStegun,NumericalRecipes}. Our computations are performed on square domains $150\times150$ with periodic boundary conditions with spatial resolution $\Delta h=0.5$, where the unit length is chosen to be the microtubular length (see Section~\ref{sec:models} for details of our dimensional units); the timestep is set to $\Delta t=0.005$. Unless explicitly stated, we set $\alpha=0.35$ and $\tau_0=0$, as discussed above. Below, we present our simulation results in composite images showing simultaneously the local density profile $\rho(\vect{r})$ (colour) and the polarisation vector field $p(\vect{r})$ (arrows), normalised by its magnitude in the globally-ordered state, Eq.\eqref{Pglob}. 

We start by examining the behaviour of the ATC model for $\Xi=0.2$ where, according to Fig.\ref{FigRhoKsi} one should expect a transition to global polar order for sufficiently high densities. For $\rho_0=2$ and $3$, there exists no instability of the homogeneous and isotropic state, and any random initial condition in our simulations quickly returns to that state. For densities above the global-instability threshold (black dotted line in Fig.\ref{FigRhoKsi}), we observe rapid formation of a globally oriented state with a large number of defects, as can be seen from Fig.\ref{FigATC02r4} for $\rho_0=5$. These defects consist of vortices, inward-pointing asters that correspond to an increase of the local density, and spatially-distributed defects of the opposite topological charge that correspond to the minima of the local density. After sufficiently long simulation times, these defects annihilate leaving behind a uniform, globally polarised state. The same behaviour persists at higher densities, the only difference {\red being that there are now} sharper density gradients around topological defects. We also observe that the typical time for all defects to annihilate grows quickly with $\rho_0$. In Fig.\ref{FigATC02r7}, for instance, we show the final snapshot of a long run for $\rho_0=12$, which continued to  coarsen over the course of the {\red whole} simulation.

\begin{figure}[t]
\subfloat[$\rho_0=5$]{
  \includegraphics[clip,width=0.5\columnwidth]{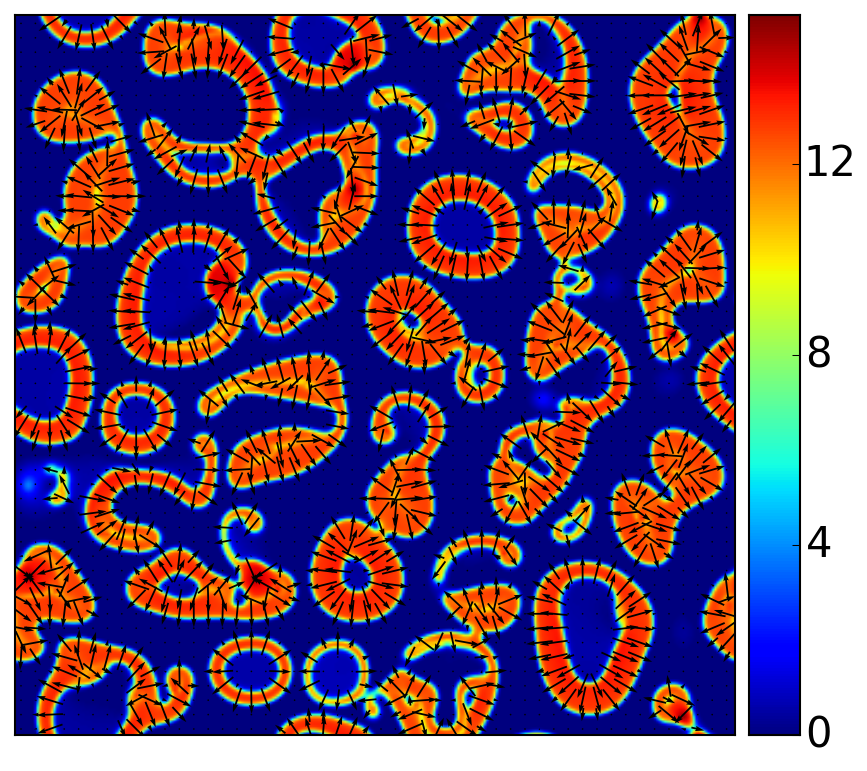}
  \label{FigATC04r4}
}
\subfloat[$\rho_0=7$]{
  \includegraphics[clip,width=0.5\columnwidth]{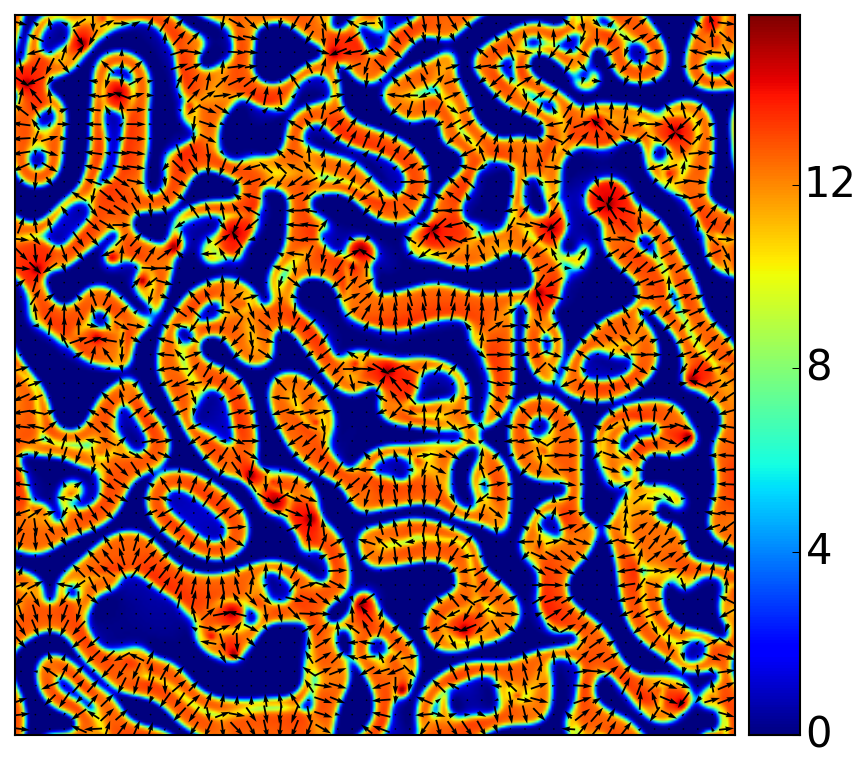}
  \label{FigATC04r5}
}

\subfloat[$\rho_0=12$]{
  \includegraphics[clip,width=0.5\columnwidth]{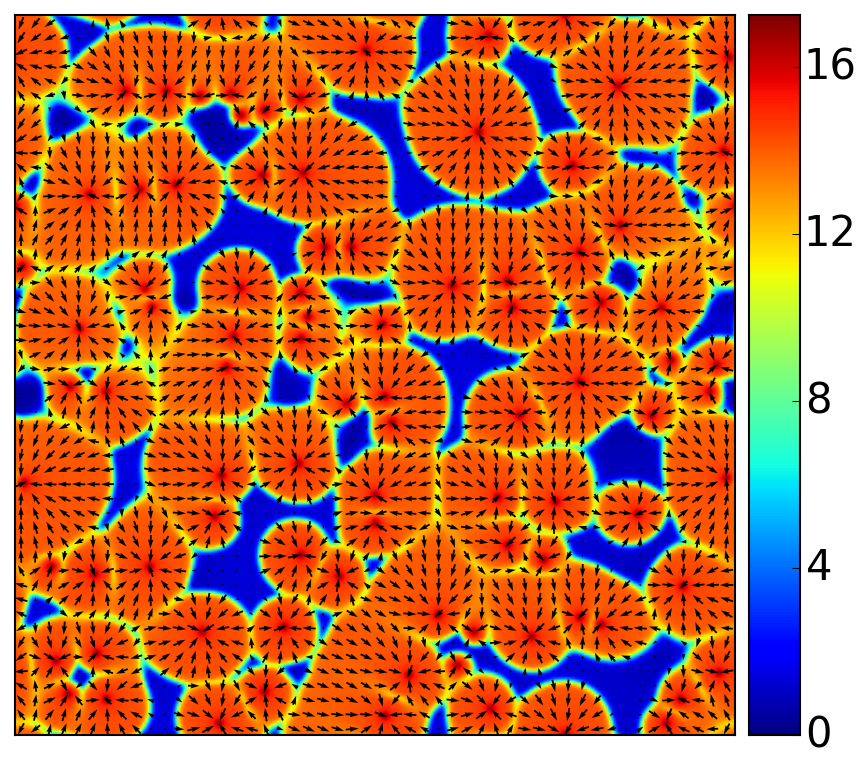}
   \label{FigATC04r8}
}
\caption{Same as Fig.\ref{FigATC02} but with $\Xi = 0.5$.}
\label{FigATC04}
\end{figure}

At $\Xi=0.5$ the behaviour of the ATC model changes considerably. According to Fig.\ref{FigRhoKsi}, as the density is increased, the bundling instability is the first {\red one} to set in. For larger densities, the bundling instability co-exists with the globally polarised state, while at yet large densities, one should again expect uniform polar order throughout the system. This scenario is supported by our direct numerical simulations.  Below the bundling instability threshold, the system always returns to the homogeneous and isotropic state. At higher densities, we observe {\red the following} dynamical structures.  For $\rho_0=5$ and $\rho_0=7$, (Figs.\ref{FigATC04r4} and \ref{FigATC04r5}, respectively), the bundling instability {\red competes} with the emergence of global order, and {\red the ensuing} high-density clusters tend to elongate to keep local polarisation aligned. Such elongated clusters often end up in yet-higher-density regions with inward-pointing asters. Even after a long time, the system does not settle into a steady-state; {\red instead its dynamics comprise} slow re-arrangements of the high-density clusters, mostly along the direction set by the local polarisation, {\red punctuated by} fast re-orientation waves that align locally the polarisation vector with the density gradient.  A similar behaviour is observed in simulations with $\rho_0=3$, which is within a narrow range of densities that are below the global instability threshold, but above the bundling instability one. In this case the system first develops clusters of high density dispersed in a low-density background until the local density inside the clusters exceeds the global instability threshold, {\red after which the} dynamics resemble its higher-density counterpart discussed above.  At yet higher density, above the bundling instability region {\red ($\rho_0=12$, see Fig.\ref{FigATC04r8}), the system does not exhibit global order as predicted by the linear stability analysis (Fig.\ref{FigRhoKsi}).} Instead it forms high-density clusters, which {\red tend to} merge into large-scale structures at very long times, see Fig.\ref{FigATC04r8}. Each cluster contains polarisation field in the form of inward-pointing asters. Perhaps, this {\red state may be viewed as an example of microphase separation, as clusters do not coarsen indefinitely but appear to reach a self-limiting size.} However, we do not know whether it survives at yet longer simulation times or in larger systems.

\begin{figure}[t]
\subfloat[$\rho_0=5$]{%
  \includegraphics[clip,width=0.48\columnwidth]{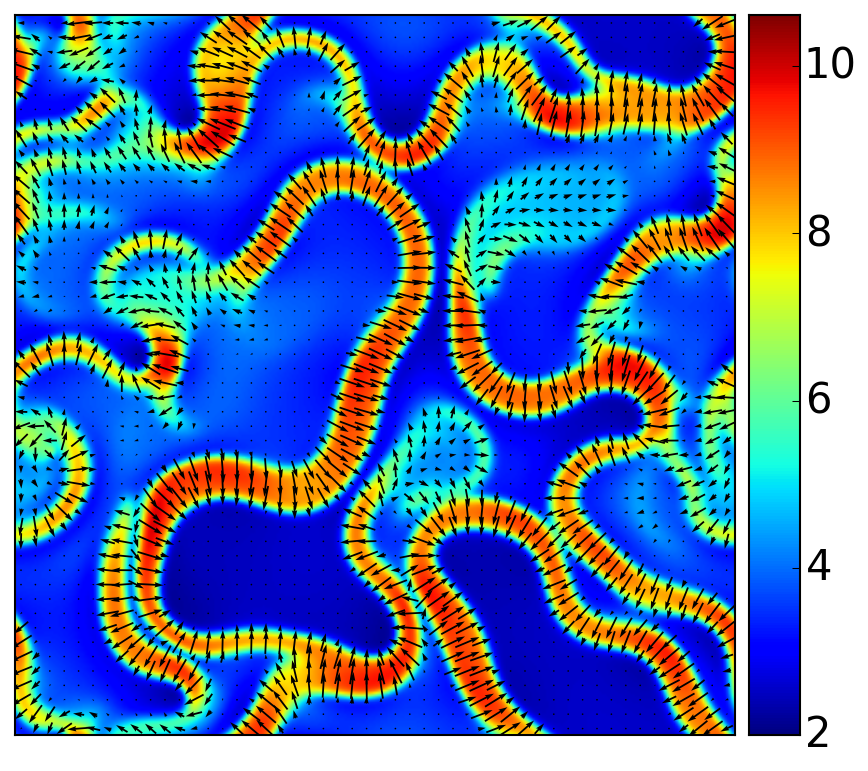}%
}
\subfloat[$\rho_0=11$]{%
  \includegraphics[clip,width=0.48\columnwidth]{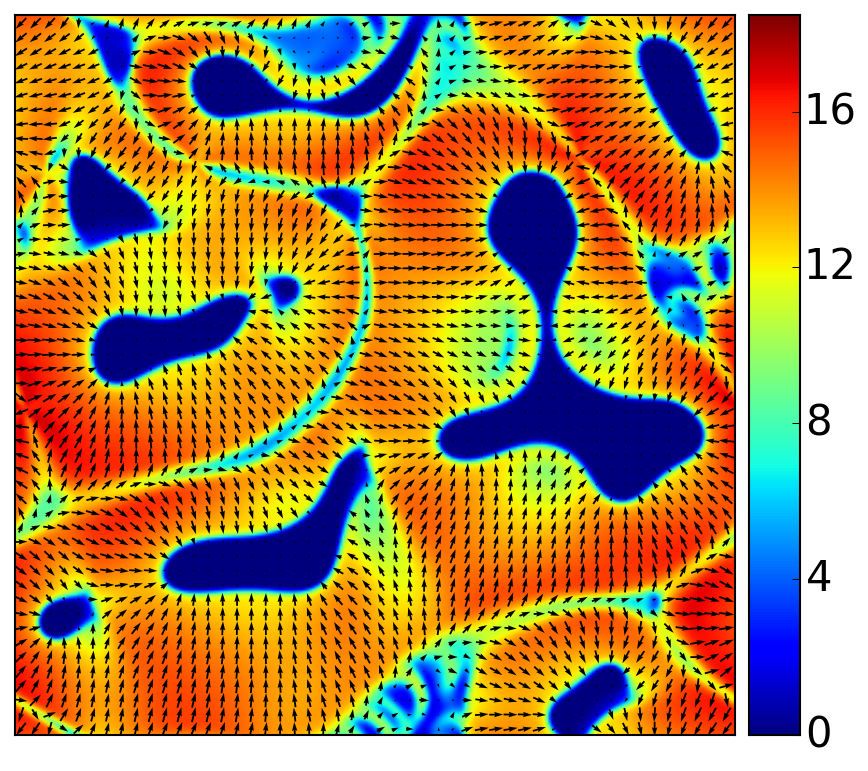}%
}
\caption{Instantaneous snapshots from the direct numerical simulations of the QC model with $\tau_0=0$, $\alpha=0.35$, and $\Xi = 0.2$.}
\label{FigQC02}
\end{figure}

\begin{figure}[t]
\subfloat[$\rho_0=4$]{%
  \includegraphics[clip,width=0.48\columnwidth]{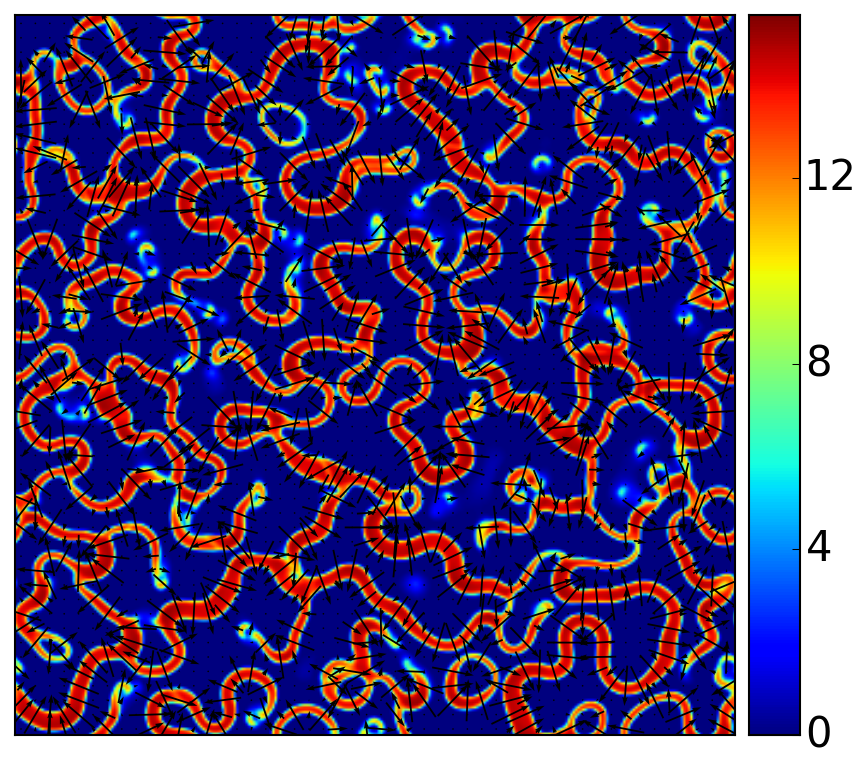}%
}
\subfloat[$\rho_0=8$]{%
  \includegraphics[clip,width=0.48\columnwidth]{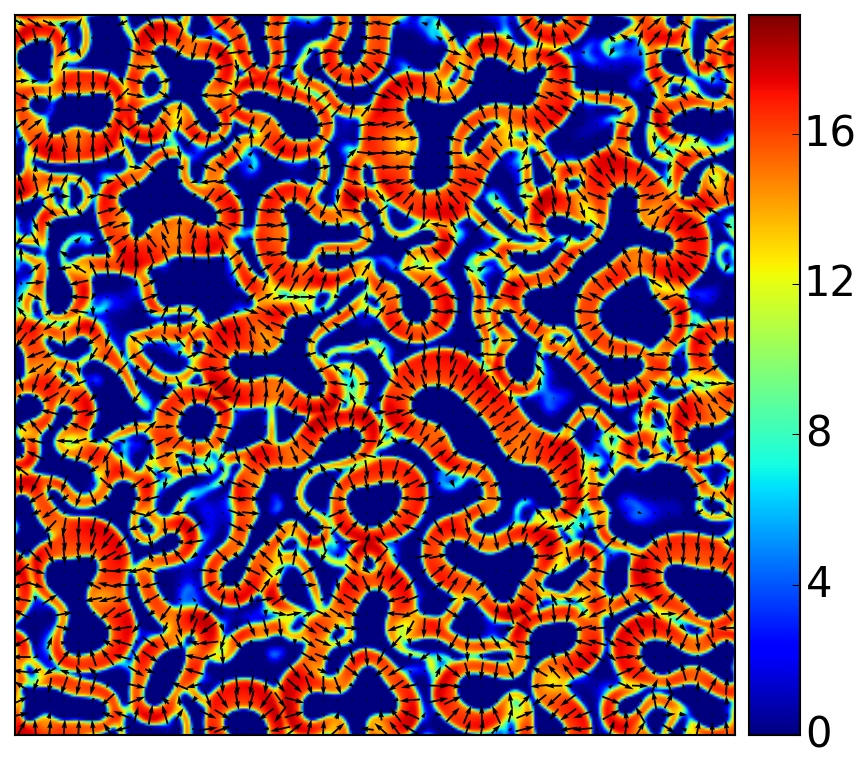}%
}

\subfloat[$\rho_0=12$]{%
  \includegraphics[clip,width=0.48\columnwidth]{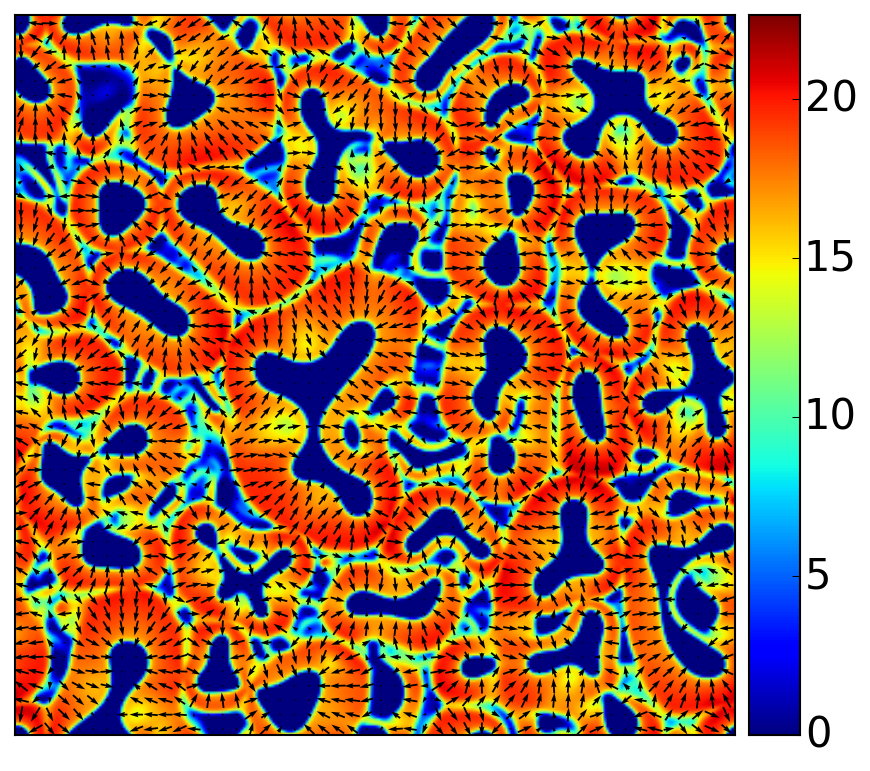}%
}
\caption{Same as Fig.\ref{FigQC02} but with $\Xi = 0.5$.}
\label{FigQC04}
\end{figure}

Now we compare these observations against the results of our direct numerical simulations of the QC model. Since the linear stability properties of the homogeneous and isotropic state are the same for both models, one might expect the QC model to exhibit a dynamical behaviour similar to the ATC one. {\red Surprisingly, the two models are instead substantially different.} As for the ATC model, {\red the cases of} $\rho_0=2$, with $\Xi=0.2$ and $0.5$, and of $\rho_0=3$, with $\Xi=0.2$, yield no instabilities, and the system returns to the homogeneous and isotropic state. Above the global instability threshold, the QC model exhibits the same type of dynamics for both $\Xi=0.2$ and $\Xi=0.5$ {\red (see Fig.\ref{FigQC02} and Fig.\ref{FigQC04}, respectively)}. Although visually these  structures appear to be similar to the ATC patterns at $\Xi=0.5$ (see Figs.\ref{FigATC04r4} and \ref{FigATC04r5}, for instance), their dynamical signatures are very different (see Supplementary Movies~\cite{SI}). While the high-density clusters of the ATC model exhibit slow, largely coarsening-type dynamics with the polarisation quickly adjusting to slowly evolving local density gradients, here the density and polarisation evolve on comparable timescales, never settle down, and appear to be chaotic for any value of $\Xi$ and $\rho_0$ in Figs.\ref{FigQC02} and \ref{FigQC04}. Even in the regions of approximately homogeneous local density, the polarisation field exhibits significant time dependence, suggesting that the globally polarised state is linearly unstable for these parameters.

\begin{figure}[t]
\includegraphics[width=1.0\columnwidth]{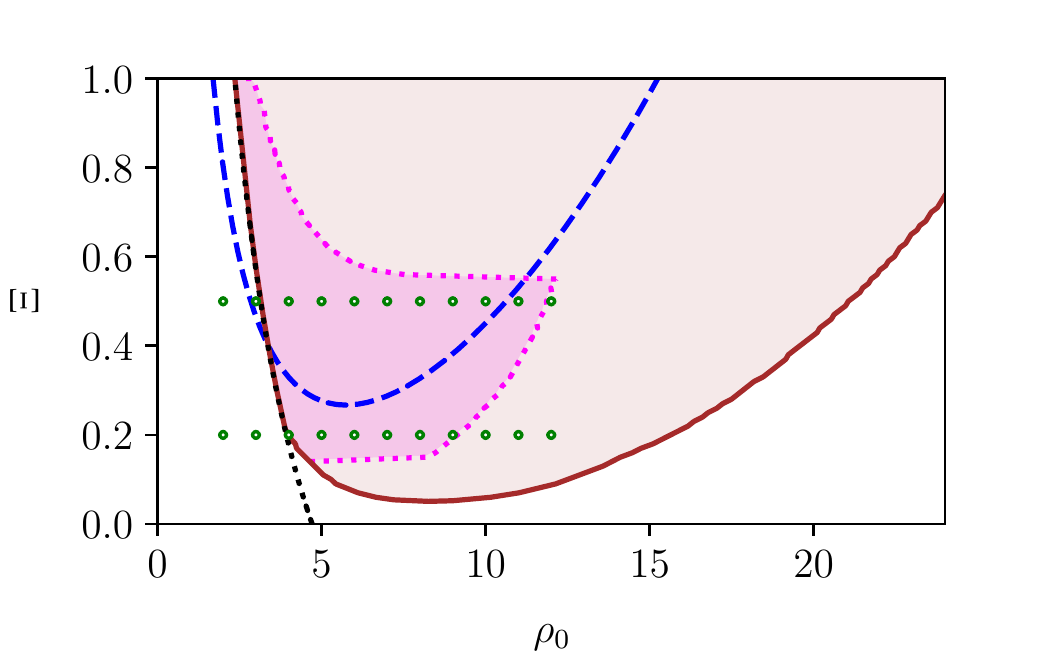}
\caption{Linear stability diagram of the QC model for $\tau_0=0$ and $\alpha = 0.35$. As in Fig.\ref{FigRhoKsi}, the dotted black line is the onset of global order, given by Eq.\eqref{rhocr}, and the dashed blue line delineates the region of the parameter space where the homogeneous and isotropic state exhibits the bundling instability (the same as the solid blue line in Fig.\ref{FigRhoKsi}). The brown solid line indicates the region where a homogeneous, globally-ordered state becomes linearly unstable. 
Inside this line we also specify the instability mode: magenta-shaded region corresponds to the density and polarisation fluctuations modulated along the direction of the global order, while the brown-shaded region corresponds to modulations both perpendicular and parallel to that direction. }
\label{FigStabDiag}
\end{figure}

To {\red validate} this statement, we performed a linear stability analysis of the globally polarised state for the ATC and QC models, see Supplemental Material for details~\cite{SI}. First, {\red this analysis confirms} that the homogeneous and isotropic state, $P=0$, of both models does not have any other instability than the bundling and global-order instabilities, discussed above. Next, we observe that while the globally polarised state is always linearly stable for the ATC model, for the QC model there is a range of parameters where it becomes unstable with respect to coupled polarisation and density fluctuations. In Fig.\ref{FigStabDiag} we plot the results of both types of linear stability analysis of the QC model. There, the black dotted line and the blue dashed line (both taken from Fig.\ref{FigRhoKsi}) correspond to the instability boundary of the globally-oriented state and the region of the bundling instability, respectively. The solid brown line marks the {\red boundary above which} the globally ordered state is linearly unstable. Additionally, within that region {\red there are two possible instability modes.} {\red The first one is characterised by a modulation in the density and polarisation {\it along}} the direction of the global order (magenta shaded region); {\red the second has} modulations {\it both perpendicular and parallel} to that direction (brown shaded region). We, therefore, speculate that {\red when there is global order (i.e., above or to the right of the dotted black line in Fig.\ref{FigStabDiag})}, the QC model exhibits three {\red types of behaviour that cannot co-exist: (i) the tendency to create global orientation with a uniform density profile, (ii) the bundling instability, and (iii) the instability of the global order.} The interaction {\red between these three instabilities is what leads} to irregular dynamics, as we see in Figs.\ref{FigQC02} and \ref{FigQC04} and Supplemental Movies \cite{SI}. 

As we can see from Fig.\ref{FigStabDiag}, for $\rho_0>3$, all our simulations (green circles) belong to the unstable region of the parameter space. We therefore performed additional simulations (not shown) of the QC model for $\Xi=0.2$ with $\rho_0=20$ and $\Xi=0.5$ with $\rho_0=25$, that both lie outside the unstable region (brown line), and confirmed the absence of chaotic-like behaviour at long times. Instead, both systems settled into a globally polarised state interlaid with topological defects, {\red similarly to the case of the ATC model}.

\begin{figure}[t]
\subfloat[ATC model]{%
  \includegraphics[width=0.53\columnwidth]{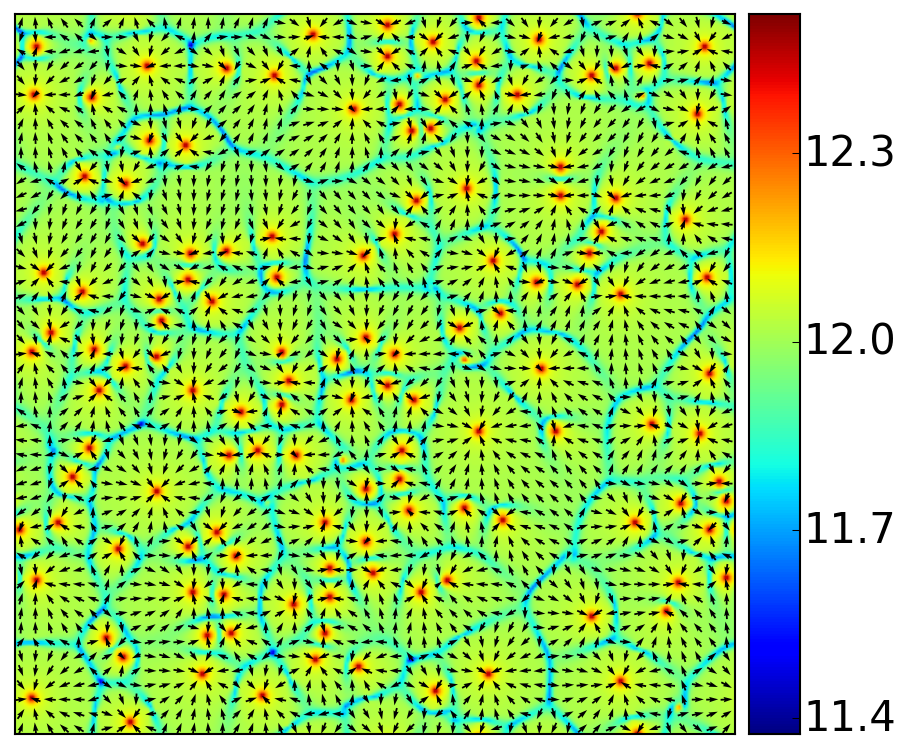}%
  \label{FigA0.5AT}
}
\subfloat[QC model]{%
  \includegraphics[width=0.5\columnwidth]{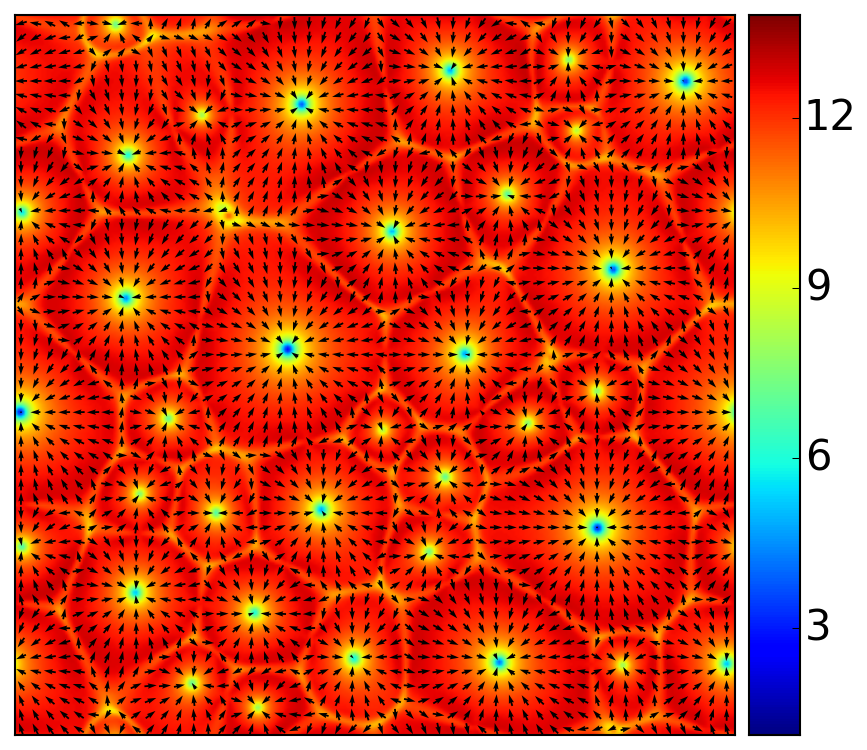}%
   \label{FigA0.5Q}

}
\caption{Comparison between the long-time dynamics of the ATC and QC models with $\tau_0=0$, $\alpha=0.6$, $\Xi = 0.5$ and $\rho_0=12$.}
\label{FigA0.5}
\end{figure}

The non-trivial dynamics presented above relies on the simultaneous existence of at least two types of instability for the same values of $\Xi$ and $\rho_0$. Fig.\ref{FigRhoAlpha} suggests that for moderate values of $\Xi$, the bundling instability only exists for small values of $\alpha$. To study the dynamics of both models outside of this regime, we set $\alpha=0.6$ and considered $\Xi=0.2$ and $\Xi=0.5$, as before. The linear stability analysis of the globally polarised state suggests that both {\red the ATC and QC} models are linearly stable in that regime, and the only instability threshold is given by Eq.\eqref{rhocr}. Our simulations confirm that both models exhibit rather simple dynamics, similar to the case of the ATC model with $\Xi=0.2$ and $\alpha=0.35$: below $\rho_{cr}$, the system returns to the homogeneous and isotropic state, while above $\rho_{cr}$, it goes through a series of long-lived topological defects before, eventually, settling into the homogeneous and isotropic state. At the highest density considered, $\rho_0=12$, the system gets trapped into a state with {\red an apparently stable (or long-lived metastable)} arrangement of topological defects (see Fig.\ref{FigA0.5}). The main difference between the two models, however, is that the inward-pointing asters of the ATC model correspond to local density enhancement, while similar topological defects in the QC model lead to local density minima.

The two situations presented above, $\alpha=0.35$ and $\alpha=0.6$, seem to comprehensively cover the behaviour of the ATC and Q models, and we have not observed any other dynamical structures besides the patterns presented above. As mentioned at the beginning of this Section, we restricted our simulations to a realistic, albeit arbitrary, case of $\tau_0=0$. Another value of $\tau_0$ would lead to a quantitative effect on the instability boundaries, while the qualitative behaviour is still the same. This is only the case for $1+a_5>0$, as Eq.\eqref{eigen_b} suggests, which is always true for $\Xi<1$. When  $1+a_5<0$, the bilaplacian terms in Eqs.\eqref{OurATrho} and \eqref{ourRhowithQ} do not result in the lengthscale selection for the bundling instability, and a yet higher-order gradient has to be added to the equations in that case.

\section{Discussion}

The main goal of this study was to {\red revisit the kinetic theory of microtubule-motor mixtures originally derived in~\cite{Aranson2006}, as well as its coarse-graining into a set of dynamical equations for (slowly-varying) density and orientation fields, Eqs.\eqref{ATrhodot} and \eqref{ATpdot}. We also studied (by linear stability analysis and direct numerical simulations) the resulting equations, and analysed the corresponding pattern formation dynamics.}

In particular, we considered the {\red validity of the effective} interaction kernel, Eq.\eqref{ATkernel}, used in~\cite{Aranson2006}. To address this issue, we developed a semi-analytical method that allowed us to calculate the interaction integrals, Eqs.\eqref{Iint} and \eqref{Jint}, exactly. We also studied the closure relationship, Eq.\eqref{adiabaticp2}, used in~\cite{Aranson2006}, and compared it to a closure method routinely used in Ginzburg-Landau-type theories of pattern formation \cite{CrossHohenberg,Ziebert2005,Peshkov2012,Peshkov2014}. We derived two dynamical systems of equations, {\red which we respectively called ATC model and QC model}, that utilise our approximation-free values of the interaction integrals, but use various closure relations. {\red While the} ATC model uses the same closure as~\cite{Aranson2006}, {\red the} QC model uses the self-consistent closure derived in Section~\ref{Q-t-c}. Together with the original equations of Aranson and Tsimring, Eqs.\eqref{ATrhodot} and \eqref{ATpdot} {\red (which we refer to as the original Aranson-Tsimring model),} these models allowed us to assess the importance of each of the assumptions mentioned above.

We used a linear stability analysis and direct numerical simulations to compare these three models. For the parameters of the {\red effective} kernel chosen by Aranson and Tsimring~\cite{Aranson2006}, the model predicts three types of behaviour: (i) the homogeneous and isotropic state for low densities, (ii) the globally-polarised state with various topological defects for intermediate densities, and (iii) the {\red bundling instability} leading to the formation of high-density clusters at high densities. Our analysis with the exact kernel demonstrated that under similar assumptions the order of the phases is different, with the bundling instability often setting in \emph{before} the globally-polarised state. Therefore, in order to {\red fully resolve the dynamics at late times}, the equations of motion should have a physical mechanism that limits the otherwise unchecked growth of the bundling instability. {\red The original Aranson-Tsimring model simply} relies on the non-linear coupling terms (i.e., terms proportional to $H$) in Eq.\eqref{ATrhodot} to cut the growth of density fluctuations -- {\red however this is a viable route only for} sufficiently large values of $H$. To cure this problem we introduced steric repulsion between the microtubular {\red bundles: this} has to be calculated up to the third virial coefficient or higher in order to provide a stabilisation mechanism that works for any density. This {\red procedure} allowed us to resolve the dynamics of our models in the region of the parameter space where the bundling and global instabilities co-exist. 
Our main conclusion here is that the usage of the exact kernel significantly alters the positions of the instability boundaries and, unless the exclusion volume parameter $\alpha$ is rather large, the bundling instability co-exists with the global order, leading to patterns absent from the {\red original Aranson-Tsimring model}~\cite{Aranson2006}. When the bundling instability is absent, the ATC model exhibits the transition to a globally-polarised state, mediated by a variety of topological defects, similar to the {\red original Aranson-Tsimring model}~\cite{Aranson2006}.

Additionally, by comparing the ATC and QC models, we concluded that the self-consistent closure employed in the latter model, changes the stability properties of the globally-polarised state in the region of the parameter space where it co-exists with the bundling instability, leading to seemingly chaotic patterns. Also, the topological defects observed for this model in the absence of the bundling instability are of rather different nature than the corresponding defects in the ATC or {\red original} Aranson-Tsimring models. 

We, therefore, conclude that out of the three sets of equations we compared, {\red the QC model more} faithfully reproduces the long-wavelength dynamics of Eq.\eqref{MEexpanded} with Eq.\eqref{exactkernel}. When either the {\red effective} kernel Eq.\eqref{ATkernel} or a closure similar to Eq.\eqref{adiabaticp2} is employed, the resulting phase diagram differs significantly from the phase diagram of the QC model. This {\red suggests that it might be of interest to analyse how the results in previous studies on microtubule-motor mixtures such as~\cite{Ziebert2007} may be affected by the use of the QC equations of motion.}

We would like to point out that there is another {\red additional remarkable} difference between the QC and the other two models: in the absence of the motor asymmetry along microtubular bundles -- i.e., for $\Xi=0$ -- the density and the polarisation equations of the Aranson-Tsimring and the ATC models decouple from each other, {\red while this is not the case in the QC model, which still} exhibits dynamical, seemingly chaotic patterns, similar to the $\Xi\ne0$ case (not shown).

Finally, we note that as the main goal of this study was to hone the techniques required to derive consistent hydrodynamic equations, we adopted {\red a simple set of} collision rules, formulated in Fig.\ref{fig:Collision_rule}, that are the basis for Eq.\eqref{MEfull}, and the expression for the interaction kernel, Eq.\eqref{exactkernel}. Detailed studies of the interactions between microtubular bundles \cite{Hentrich2010,Henkin2014,Cross2014,Hilitski2015} suggest that these assumptions {\red are unlikely to be fully} realistic, and {\red will} require refinement. Additionally, there is a need to understand the role that potential microtubular self-propulsion, discussed by Liverpool, Marchetti and co-workers~\cite{Liverpool2003,Liverpool2005,Ahmadi2006,Marchetti2013}, might play in the dynamics of microtubules-molecular motor mixtures. We plan to address some of these questions in our future work.

\section{Acknowledgement}

Discussions with Igor Aranson and Lev Tsimring are kindly acknowledged.
AG acknowledges funding from the Biotechnology and Biological Sciences Research Council of UK (BB/P01190X, BB/P006507). DM acknowledges support from ERC CoG 648050 (THREEDCELLPHYSICS). Additional simulation movies and further research outputs generated for this project can be found at http://dx.doi.org/10.7488/ds/2246.

\end{document}